\documentclass[manuscript]{acmart}

\AtBeginDocument{%
  \providecommand\BibTeX{{%
    \normalfont B\kern-0.5em{\scshape i\kern-0.25em b}\kern-0.8em\TeX}}}

\setcopyright{acmlicensed}
\copyrightyear{2024}
\acmYear{2024}
\acmDOI{XXXXXXX.XXXXXXX}

\begin{document}

%%
%% The "title" command has an optional parameter,
%% allowing the author to define a "short title" to be used in page headers.
%\title{Opportunities \& Challenges for Responsible AI Design and Integration in Radiology Workflows: A Case Study of Automatic Feeding Tube Qualification}
\title{Challenges for Responsible AI Design and Workflow Integration in Healthcare: A Case Study of Automatic Feeding Tube Qualification in Radiology}
%\title{How AI Design \& Integration Choices in Radiology Workflows Implicate AI Acceptance and Risks: A Case Study of Automatic Feeding Tube Qualification}
%\title{Driving forward AI Innovation}
%\title{Understanding Opportunities \& Challenges for  AI Design and Integration into Radiology Workflows: A Case Study of Automatic Feeding Tube Qualification}
%\title{Understanding Opportunities \& Challenges for Responsible AI Design in Radiology: A Case Study of Automatic Feeding Tube Qualification}
%%
%% The "author" command and its associated commands are used to define
%% the authors and their affiliations.
%% Of note is the shared affiliation of the first two authors, and the
%% "authornote" and "authornotemark" commands
%% used to denote shared contribution to the research.
\author{Anja Thieme}
%\authornote{Both authors contributed equally to this research.}
\email{anthie@microsoft.com}
%\orcid{0000-0002-9639-5531}
\affiliation{%
  \institution{Microsoft Health Futures}
  \streetaddress{21 Station Road}
  \city{Cambridge}
  \state{Cambridgeshire}
  \country{UK}
  \postcode{CB1 2FB}
}

\author{Abhijith Rajamohan}
%\email{abhijith.rajamohan@nhs.net}
%\orcid{}
\author{Benjamin Cooper}
%\email{ben.cooper2@nhs.net}
%\orcid{}
\author{Heather Groombridge}
%\email{heather.groombridge@nhs.net}
%\orcid{}
\author{Robert Simister}
%\email{robert.simister@nhs.net }
%\orcid{}
\author{Barney Wong}
%\email{barney.wong@nhs.net}
%\orcid{}
\affiliation{%
  \institution{UCLH NHS Foundation Trust}
  \country{UK}
}
\author{Nicholas Woznitza}
%\email{n.woznitza@ucl.ac.uk}
\affiliation{%
 \institution{UCLH NHS Foundation Trust}
\country{UK}
}
\affiliation{%
 \institution{Canterbury Christ Church University}
\country{UK}
}

\author{Mark Ames Pinnock}
%\email{mark.pinnock.18@ucl.ac.uk}
%\orcid{}
\affiliation{%
  \institution{University College London}
  \country{UK}
}

\author{Maria Teodora Wetscherek}
%\email{teo.wetscherek@nhs.net}
%\orcid{0000-0003-2924-7587}
\affiliation{%
  \institution{Cambridge University Hospitals NHS Foundation Trust}
  \country{UK}
}

\author{Cecily Morrison}
%\orcid{0000-0001-5013-3715}
%email{cecilym@microsoft.com}
\affiliation{%
  \institution{Microsoft Research Cambridge}
  \country{UK}
}

\author{Hannah Richardson}
%\email{hamurfet@microsoft.com}
\author{Fernando P\'erez-Garc\'ia}
%\email{fernando.perezgarcia@microsoft.com}
%\orcid{0000-0001-9090-3024}
\author{Stephanie L. Hyland}
%\email{sthyland@microsoft.com}
\author{Shruthi Bannur}
%\email{Shruthi.Bannur@microsoft.com}
\author{Daniel C. Castro}
%\email{dacoelh@microsoft.com}
%\orcid{0000-0002-6829-7045}
\author{Kenza Bouzid}
%\email{t-kbouzid@microsoft.com}
\author{Anton Schwaighofer}
%\email{antonsc@microsoft.com}
\author{Mercy Ranjit}
%\email{mercy.ranjit@microsoft.com}
\author{Harshita Sharma}
%\email{harshita.sharma@microsoft.com}

\affiliation{%
  \institution{Microsoft Health Futures}
  \country{UK}
}

\author{Matthew P. Lungren} 
%\orcid{}
%\email{mlungren@microsoft.com}
\affiliation{%
  \institution{Microsoft Health Futures}
  \country{US}
}
\affiliation{%
  \institution{University of California}
  \country{US}
}
\affiliation{%
  \institution{Stanford University}
  \country{US}
}

\author{Ozan Oktay}
%\email{ozan.oktay@microsoft.com}
\author{Javier Alvarez-Valle}
%\email{jaalavare@microsoft.com}
\author{Aditya Nori}
%\email{adityan@microsoft.com}
\affiliation{%
  \institution{Microsoft Health Futures}
  \country{UK}
}

\author{Stephen Harris}
%\email{s.harris@nhs.net}
%\orcid{}
\author{Joseph Jacob}
%\authornotemark[1]
%\email{j.jacob@ucl.ac.uk}
\affiliation{%
  \institution{University College London}
  \country{UK}
}

%%
%% By default, the full list of authors will be used in the page
%% headers. Often, this list is too long, and will overlap
%% other information printed in the page headers. This command allows
%% the author to define a more concise list
%% of authors' names for this purpose.
\renewcommand{\shortauthors}{Thieme, et al.}

%%
%% The abstract is a short summary of the work to be presented in the
%% article.
\begin{abstract}
Nasogastric tubes (NGTs) are feeding tubes that are inserted through the nose into the stomach to deliver nutrition or medication. If not placed correctly, they can cause serious harm, even death to patients. Recent AI developments demonstrate the feasibility of robustly detecting NGT placement from Chest X-ray images to reduce risks of sub-optimally or critically placed NGTs being missed or delayed in their detection, but gaps remain in clinical practice integration. In this study, we present a human-centered approach to the problem and describe insights derived following contextual inquiry and in-depth interviews with 15 clinical stakeholders. The interviews helped understand challenges in existing workflows, and how best to align technical capabilities with user needs and expectations. We discovered the trade-offs and complexities that need consideration when choosing suitable workflow stages, target users, and design configurations for different AI proposals. We explored how to balance AI benefits and risks for healthcare staff and patients within broader organizational and medical-legal constraints. We also identified data issues related to edge cases and data biases that affect model training and evaluation; how data documentation practices influence data preparation and labelling; and how to measure relevant AI outcomes reliably in future evaluations. We discuss how our work informs design and development of AI applications that are clinically useful, ethical, and acceptable in real-world healthcare services.
%Chest X-rays (CXRs) are one of the most common imaging modalities used in clinical practice [REF], yet their interpretation can be challenging and time-consuming, especially in busy and resource-limited settings. Responding to a need for automated tools that can assist radiologists and clinicians in detecting and localizing medical lines and tubes on CXRs, ...
\end{abstract}

%%
%% The code below is generated by the tool at http://dl.acm.org/ccs.cfm.
%% Please copy and paste the code instead of the example below.
%%


%%
%% Keywords. The author(s) should pick words that accurately describe
%% the work being presented. Separate the keywords with commas.

%\received{XX Month 2024}
%\received[revised]{XX Month 2024}
%\received[accepted]{XX Month 2024}

%%
%% This command processes the author and affiliation and title
%% information and builds the first part of the formatted document.
\maketitle

\section{Introduction}
Artificial Intelligence (AI) is increasingly gaining recognition as an important application in radiology~\cite{filice2020case, huang2023artificial, huang2023generative, patel2019human, strohm2020implementation}. AI has been applied to detect and diagnose important clinical findings (e.g., ~\cite{hurt2020augmenting, jang2020deep, moore2023prevalence, wang2022efficient}, or to analyse anatomical structures on medical images~\cite{gordienko2019deep, harrison2022machine}. Latest advances in foundation models (FMs)~\cite{bommasani2021opportunities}, which are powerful, general-purpose models that can be adapted to various healthcare and radiology-specific tasks, suggest even greater potential for AI to revolutionize clinical practice~\cite{brown2020language, moor2023foundation, singhal2023large, thieme2023foundation, wojcik2022foundation} as demonstrated in tasks such as medical knowledge extraction~\cite{preston2023toward}, clinical text modification (e.g., ~\cite{jeblick2022chatgpt, krishna2020generating, NablaCopilot_2023, nuanceNuanceMicrosoft}) and new forms of medical question-answering 
~\cite{lee2023benefits, singhal2023towards}. Recent approaches towards \textit{multimodal} FMs, that integrate medical images alongside text representations (e.g.,  BioVIL(-T)~\cite{bannur2023learning, boecking2022making}, ELIXR~\cite{xu2023elixr},  MAIRA-1~\cite{Hyland2023maira}, or Med-PaLM M~\cite{tu2023towards}), further expand the scope of possible, innovative use cases. For AI assisted radiology work, this includes capabilities to: automatically generate a radiology report from a medical image (e.g., ~\cite{huang2023generative, Hyland2023maira, yu2023evaluating}); answer questions about a radiology image using text queries (cf. ~\cite{ xu2023elixr}); or detect errors in a radiology report text by comparing it with the image~\cite{yildirim2024multimodal}.
%"The enthusiasm with which data scientists and predictive analytics companies—from large, well-established companies to start ups—have embraced the application of AI in health care has resulted in a plethora of algorithms and new commercial products. Intense pressure is being placed on health care systems to implement them."~\cite{lindsell2020action}

Despite all the excitement and remarkable progress in AI research and development in recent times, the successful translation of such technical innovations into clinical practice remains challenging~\cite{baxter2020barriers, beede2020human, galsgaard2022artificial, ontika2023pairads, osman2021realizing, verma2021improving, yu2018artificial, tulk2022inclusion, strohm2020implementation}. In what has been described by Andersen et al.~ \cite{osman2021realizing} as “a race for getting the technology right before exposing human end-users to new promising AI tools”; many researchers warn against the dangers of developing AI in isolation~\cite{miller2017explainable} – without considering the specific needs of the intended users and the downstream implications of the technology~\cite{liao2022connecting}. This currently leaves a significant gap between compelling technical proofs-of-concept or lab experiments and larger ambitions of integrating and deploying AI-enabled systems successfully in routine clinical care~\cite{coiera2019last, kelly2019key, yu2018artificial}.

Some of the challenges for real-world healthcare integration arise from a lack of trust in the ‘black-box’ nature of advanced AI outputs~\cite{rudin2019stop, zajkac2023clinician}; and the difficulties for designers and healthcare experts to appropriately understand and productively work with new AI capabilities~\cite{cai2019hello, yildirim2023creating,yang2019unremarkable}. There is also uncertainty about the value that AI applications bring to clinical practice. This is evident where systems are perceived to not provide useful information~\cite{barda2020qualitative, petitgand2020investigating}, or where such information cannot, or is not actioned~\cite{ginestra2019clinician} -- reducing its usefulness. Alongside this, there are increasing demands for clinical effectiveness trials (cf.~\cite{galsgaard2022artificial}). There are also requirements for new AI solutions to align within context-specific patterns of care~\cite{petitgand2020investigating}; and to account for the disruptions they can create on existing workflows and social relationships that often characterize collaborative healthcare delivery~\cite{Elish2020Repairing}; as well as many broader organizational, ethical and regulatory issues that affect the adoption and use of AI systems in healthcare (cf. ~\cite{strohm2020implementation}) . Therefore, designing effective human-AI interactions within the clinical domain presents a complex socio-technical problem that requires a holistic, multidisciplinary approach that deeply engages with the real-world problems and contexts that AI solutions aim to solve~\cite{Elish2020Repairing, osman2021realizing}; carefully matching technical capabilities with user needs~\cite{liao2022connecting, thieme2023foundation}.

The work presented in this paper builds on recent human-centred healthcare AI research (e.g.,~\cite{beede2020human, burgess2023healthcare, calisto2021introduction,gu2023augmenting, jacobs2021designing, matthiesen2021clinician, ontika2022exploring, sendak2020human, yang2019unremarkable, yang2023harnessing}) that studies clinical workflows and corresponding AI integration challenges~\cite{beede2020human, burgess2023healthcare, calisto2021introduction, yang2023harnessing}; and provides insights into AI design~\cite{burgess2023healthcare, hirsch2018s, yang2019unremarkable} for configuring effective human-AI interactions~\cite{gu2023augmenting, thieme2023designing}. Starting as early as problem identification and ideation stages~\cite{ismail2018bridging, kross2021orienting, yildirim2023investigating, wilcox2023ai, cai2021onboarding}, our research  explores the opportunities and challenges of developing and integrating AI within an end-to-end radiology imaging workflow such that it can achieve \textit{clinical utility}. The aim is to develop an in-depth understanding of existing clinical workflows and data documentation practices as well as derive key insights and requirements that can meaningfully guide AI development (e.g., assist with data labels, edge case understanding, or evaluation metrics). More specifically, we seek to extract more clearly the \textit{interrelations}~\cite{berg2003ict, zajkac2023clinician} between AI data work or systems and corresponding clinical needs, workflows, and broader AI acceptance and organizational considerations. In this regard, our human-centred approach focuses on the specific use case of AI assisting the interpretation of Chest X-ray images (CXRs) to verify the correct placement of nasogastric feeding tubes in intensive care patients with the goal of improving workflow efficiency and patient safety.

\subsection{Use Case: Nasogastric Feeding Tube Placement Qualification on Chest X-ray Images}
A nasogastric tube (NGT) is a thin tube that is inserted via the nose and passed into the stomach. It is used for short- to medium-term nutritional support, medication administration or aspiration of stomach contents~\cite{NHSKentCommunityHealth_2022}. NGTs are amongst the most commonly used medical lines in critically ill patients in intensive care units (ICU) and emergency departments for life-supporting purposes~\cite{yi2020computer}, and high-dependency units and departments where patients require nutritional support (i.e., Stroke). Due to increases in the number of hospitalized patients, it is estimated that approximately 10 million NGTs are used annually in Europe, 1 million of which are used in the UK (approx. 1.2 million in the US)~\cite{torsy2020accuracy}. Previous research highlights a variety of complications associated with NGT placement~\cite{o2021emergency, zhang2021malposition} especially bronchial insertion (Figure~\ref{fig:NGTcases}B), which can potentially result in aspiration of feeds and pneumothorax~\cite{gimenes2019nasogastric}. These can increase time spent in intensive care, treatment costs~\cite{aryal2021identifying}, and patient morbidity and mortality, highlighting the importance of feeding tubes being placed properly and used safely~\cite{yi2020computer}. Yet, clinical studies demonstrate that up to 3\% of NGTs are reported as misplaced within the airways, which causes complications in up to 40\% of these cases~\cite{koopmann2011team}. Seeking to reduce risks of sub-optimally or critically placed NGTs being missed or delayed in their detection, initial AI developments demonstrate the feasibility of robustly detecting NGT placement from CXR images~\cite{drozdov2023artificial, shah2021machine}. Despite great technical advances, gaps remain in understanding how best to design; practically integrate; and responsibly use any such models as part of clinical workflows; as well as how best to evaluate the effectiveness of any prospective AI application. We examine these issues in the context of a UK hospital ICU, with some comparisons to Stroke care, which is detailed in Section 4.

\subsection{Research Questions \& Contributions}
Our work presents a rare example of an in-depth case study that engages early in the AI design process with the end-to-end workflow and concrete use context of CXR-based NGT placement verification with key domain stakeholders. The study aim is to understand unique opportunities and challenges for creating clinically useful AI applications. Specifically, we ask: (1) what are the right types of applications; (2) how to effectively and responsibly design those from a human-centred perspective; and (3) how can insights into the specific use context usefully guide AI development and evaluation? Against this backdrop, our work makes three main contributions:
\begin{enumerate}
    \item \textit{We surface complex interrelations between human, technical and organizational factors that determine perceived AI utility and successful adoption}; and we propose the systematic mapping of identified factors as a tool to clarify important benefit-risk and feasibility trade-offs. 
    \item \textit{We propose ‘Human-Process Integration of AI’ as an approach to future AI development to foster AI acceptance}. We argue that AI should not be seen as a separate entity that needs human ‘verification’, but framed as part of existing human (safeguarding) processes of information review, guideline adherence, and patient concern.
    %Whilst many human-AI uses require humans to ‘verify the AI’, we suggests a closer positioning of AI within existing human (safeguarding) processes by framing AI outputs as part of important information review, guideline adherence, and patient safety considerations.
    %Moving away from ‘human-in-the-loop AI’ concepts that require humans to ‘verify the AI’, it suggests a closer positioning of AI within existing human (safeguarding) processes by (i) focusing on lower-level (expert) AI analysis; and (ii) by framing AI outputs as part of important information review, guideline adherence, and patient safety considerations.}
    \item \textit{We extract key insights into real-world data availability and data production practices, and discuss their implications for AI development} (e.g., dataset curation, model training, and outcome evaluations).
     \end{enumerate}

\section{Related Work}
We begin with a concise summary of: (i) existing AI approaches for automatically detecting and localizing medical lines and tubes on CXRs; and (ii) the relevant literature on human-centered AI research in healthcare and radiology. 

\subsection{AI \& Machine Learning for Automated Detection of Medical Lines and Tubes on CXRs}
In recent years, we witness a growth of AI research and development in the automatic detection and localization of medical lines and tubes on CXRs, seeking to help prioritize and shorten turn-around times especially for critical cases (e.g., ~\cite{lee2018deep, rungta2021detection, singh2019assessment, yi2020computer}) and thereby improve the effectiveness of clinician workflows and patient safety~\cite{elaanba2021automatic, singh2019assessment, yu2020detection}. The majority of existing works is focused on detecting one specific tube type, most commonly \textit{central venous catheters (CVC)}~\cite{lee2018deep, shah2021machine, sirazitdinov2021landmark,subramanian2019automated, yu2020detection}, which are thin tubes inserted via the patient’s veins to draw blood and give treatments~\cite{sirazitdinov2021landmark}; and also \textit{endotracheal tubes (ETT)}~\cite{kao2015automated}, which are airway tubes to assist in lung ventilation. Given that critically ill patients often have multiple lines and tubes inserted (i.e., patients are intubated for air ventilation and fed via a feeding tube), many studies explored the differentiation of multiple tube types on a CXR image~\cite{abbas2022automatic, aryal2021identifying, borvornvitchotikarn2022pre, elaanba2021automatic, henderson2021automatic, khan2021early, rungta2021detection}. 

Only few studies to date specifically address the placement of feeding tubes~\cite{drozdov2023artificial, singh2019assessment}. Most notably, Drozdov et al. ~\cite{drozdov2023artificial} report the development and evaluation of a deep learning (DL) approach for NGT misplacement detection. Their model achieves high performance on various NGT classification tasks (e.g.,AUC of 0.98 for lung malplacement). The authors are also amongst few (e.g.,~\cite{seah2021effect}) who study how CXR AI can assist clinicians in critical tube findings detection and enhance clinical decision-making. Their study with five clinicians reviewing 335 CXR images with and without AI revealed an increased of overall accuracy in decision to feed from 69\% (unaided) to 78\% (with AI), suggesting greater clinician confidence in decision making, and the potential for AI to reduce NGT misplacement complications. 
%Amongst the few works that study the use of chest x-ray AI for critical findings detection with clinicians for improving clinical decision-making, Seah et al. [REF] report the findings of an evaluative study with 20 radiologist who reviewed CXR cases with and without the assistance of a deep learning model that classified clinical findings. They showed how radiologists classification accuracy significantly improved with the DL model on 102 out of 127 clinical findings (80%); demonstrating the potential of AI to improve chest x-ray interpretation across a large breadth of clinical practice.

More generally, for tube placement detection, we find a variety of ML approaches applied - mostly to CXR image analysis; and in rare cases to radiology reports (cf.~\cite{shah2021machine}). These approaches seek to assist in tasks such as:  (i) detecting the \textit{existence of a tube} on a CXR~\cite{henderson2021automatic,kao2015automated,subramanian2019automated}; (ii) classifying the \textit{tube type} present~\cite{abbas2022automatic, henderson2021automatic, khan2021early, subramanian2019automated}; (iii) identifying or classifying \textit{tube tip position} ~\cite{lee2018deep,kao2015automated,shah2021machine, yu2020detection} and \textit{relevant landmarks}~\cite{sirazitdinov2021landmark}; (iv) and classifying the \textit{accuracy} (i.e., normal, borderline, abnormal ~\cite{abbas2022automatic,aryal2021identifying, borvornvitchotikarn2022pre,drozdov2023artificial, khan2021early, rungta2021detection}) or \textit{criticality of the tubes placement} (e.g., critical vs. non-critical~\cite{seah2021effect, singh2019assessment}). Most of the datasets used for analysis are either self-curated; or derived from much larger publicly available datasets like: MIMIC~\cite{johnson2019mimic}, NIH ChestX-ray14~\cite{wang2017chestx} and RANZCR CLiP~\cite{tang2021clip}. See Table \ref{Table: TubeStudies} in the Appendix for an overview of these studies, including datasets used, and reported AI performance outcomes. In general, across these studies, tube misplacement classification performance is high, with many reporting accuracies of 90-95\%, or more.  

In terms of more commercially oriented developments, \textit{Qure.ai} reported receiving FDA approval\footnote{https://qure.ai/news\_press\_coverages/qure-ais-breathing-tube-placement-ai-technology-receives-fda-clearance/} for ML confirmed placement of breathing tubes. 
%Detecting the carina; breathing tube tip; and distance between those two structures, they reported ML performance on a sample of 162 studies that showed an Absolute Error in distance between breathing tube tip and carina of 1.98 mm (SD = 1.41). 
The company \textit{annalise.ai} employed a deep learning (DL) model on a large scale of CXRs (over 800,000 images) with clinician curated labels for a wide range of clinical findings; their model was able to robustly predict suboptimally placed NGTs with high AUC (0.984) alongside other catheter types such as central lines, ETTs and pulmonary arterial catheters~\cite{seah2021effect}. Lastly, based on the research by Drozdov et al. ~\cite{drozdov2023artificial} reported above, \textit{Bering Ltd} recently received UK CA marking\footnote{https://icaird.com/2022/icaird-partner-bering-achieves-ukca-mark-for-ai-supported-chest-x-ray-classification/} for BraveNGT, an AI software that detects NGT malposition on CXRs to provide effective decision support for clinicians whether feeding in patients can be safely performed.

All these works suggest growing research and commercial interest and increasing promise of utilizing AI capabilities in CXR analysis to provide useful insights to lines and tubes (mis)placement qualification. Simultaneously, it demonstrates the need to move from technical solutions and clinical proof-of-concepts to understanding AI system deployment and how design choices implicate both utility and risks. 

\subsection{Human-Centered AI Research \& Design in Healthcare and Radiology}
AI in healthcare is a complex and challenging domain that requires human-centered research and design approaches~\cite{beede2020human, jacobs2021designing, zajkac2023clinician}. Below, we first provide a brief summary of common challenges for clinical AI development, before extending into the domain of medical imaging and radiology more specifically.
\subsubsection{Challenges for Clinical AI} 
Developing AI solutions for healthcare involves many technical, ethical, and social issues that need to be carefully considered and addressed. Some of these issues include the quality of healthcare data used to train, adapt or evaluate AI models; and related concerns about how patients can control, consent or opt out of data uses; and how their data privacy and security can be ensured~\cite{thieme2020machine, wilcox2023ai}. Research also engages with AI acceptance and adoption challenges~\cite{yang2023harnessing, jacobs2021designing, matthiesen2021clinician, sendak2020human}. For example, to foster trust and confidence in AI, the field of eXplainable AI (XAI) aims to make AI more transparent and understandable through various methods of explanation and feedback that enable clinicians to contest~\cite{hirsch2017designing}, verify~\cite{yang2023harnessing}, and learn from AI outputs~\cite{cai2019hello}; seeking to configure effective human-AI collaboration~\cite{thieme2020interpretability}. Moreover, AI in healthcare must address the potential risks of inequality and discrimination that may arise from biased or unfair data algorithms~\cite{bender2021dangers, chien2022multi, obermeyer2019dissecting, thieme2023foundation,zack2023coding}; and the need for rigorous evaluation frameworks that assess and monitor AI performance in real-world settings~\cite{liang2022holistic}. Moreover, there are many broader organizational, social, ethical and regulatory implications (cf. ~\cite{strohm2020implementation}) that raise questions about who is accountable and responsible for any AI-assisted decisions in healthcare spanning individual users, healthcare institutions, and insurance providers (e.g., ~\cite{gilbert2023large, petersen2022responsible, procter2023holding}). As a diverse and heterogeneous domain that encompasses different specialties, settings, workflows, and stakeholders, AI solutions for healthcare therefore require tailoring to the specific needs of the intended users and beneficiaries. Against this backdrop, our work seeks to develop a deeper understanding of existing work practices and the clinical problem space that surrounding NGT placement verification within an ICU setting, with the aim to better understand the specific use context, design requirements, and potential impacts of any prospective AI intervention.

\subsubsection{Human-centered AI in Radiology}
Human-centered AI research in medical imaging spans investigations in the fields of ophthalmology~\cite{bach2023if, beede2020human}, pathology~\cite{cai2019human, gu2023augmenting, gu2023improving, lindvall2021rapid}, and radiology~\cite{atad2022chexplaining, bernstein2023can, calisto2021introduction, calisto2022breastscreening, calisto2023assertiveness, ontika2023pairads, verma2021improving, xie2020chexplain}. In pathology, for instance, Cai et al.~\cite{cai2019hello, cai2019human} presented their user research and development process for the \textit{SMILY} prototype, a prediction tool for prostate cancer diagnosis. They showed how enabling users to interactively refine the predictions improved the clinical utility of their tool and user trust in the algorithm. Lindvall et al.~\cite{lindvall2021rapid} designed \textit{Rapid Assisted Visual Search}, a human-AI interface to help pathologists assess colorectal cancer. Conducting an evaluation with six pathologists, they demonstrated how their interface reduced pathologists' search time for regional lymph nodes with signs of metastasis.  

In radiology more specifically, AI research so far mainly focused on making AI outputs more understandable to domain experts~\cite{atad2022chexplaining, calisto2021introduction, calisto2022breastscreening, ontika2023pairads}. For example, Atad et al.~\cite{atad2022chexplaining} used counterfactuals to explain the AI diagnosis of CXR findings (e.g., cardiomegaly) by highlighting what feature changes in the image would lead the model to give a different outcome. Ontika et al.~\cite{ontika2023pairads} proposed a hybrid AI system for multiparametric MRI to help prostate cancer diagnosis; using visualizations to facilitate human comprehension of AI generated outputs. The authors conducted contextual inquiries and interviewed  five radiologists to better understand how to create ``an impactful human-AI collaborative environment in radiology'' (p. 395). Their work revealed the variability of medical imaging interpretation, workflow differences across radiology centers, and potential for automating tedious and repetitive tasks.

A growing area of research is also the evaluation of AI radiology models and the study of their clinical impact. More technically-oriented investigations focus on new approaches to systematically evaluating the accuracy and suitability of model outputs, and identifying relevant radiology-specific metrics (e.g., assessing factual completeness and consistency of AI generated radiology text~\cite{miura2020improving}); often these involve domain experts in reviewing and categorizing AI errors~\cite{Hyland2023maira, yu2023evaluating}. Other studies examine the effects of AI use on clinicians, demonstrating improved radiologists classification accuracy with AI~\cite{seah2021effect}, and reduced diagnosis time and error rates of clinicians (cf.  AI assisted \textit{BreastScreening} for breast cancer diagnosis ~\cite{calisto2021introduction, calisto2022breastscreening}). Some studies also explored how different ways of presenting AI outputs to clinicians (e.g., via an assertive or non-assertive communication style) can reduce medical errors~\cite{calisto2023assertiveness}. For instance, Bernstein et al.~\cite{bernstein2023can} showed how incorrect AI outputs can bias radiologists to make incorrect follow-up decisions when they were correct without AI. However, adding a bounding box on the region of interest (e.g., to verify AI results) reduced human errors. 

Lastly, few studies investigate radiology workflows or focus on understanding current needs of radiologists in their daily practice~\cite{ontika2022exploring, verma2021improving, xie2020chexplain, yildirim2024multimodal}. An exception is work by Xie et al., who conducted an early phase needfinding and design study that included a survey, low-fidelity prototype design, and high-fidelity evaluation to identify opportunities for AI-assistance in radiology X-ray work~\cite{xie2020chexplain}. Exploring the barriers to AI adoption in radiology imaging, Verma et al.'s also reported an interview study with seven imaging experts in oncology, which revealed rich insights into clinicians' concerns about black-box models, small training datasets, and quality control in training data~\cite{verma2021improving}.

Crucially, these works (e.g.,~\cite{ bernstein2023can, calisto2023assertiveness}) evidence that AI implementation choices affect human-AI performance; emphasizing the need to further investigate AI design and integration challenges and opportunities. We extend this line of work by investigating the end-to-end workflow in verifying feeding tube placement via CXRs to learn about challenges and identify design requirements for clinically relevant, responsible AI development.

\section{Study Method}
Aiming to clarify opportunities and challenges for AI assisted NGT placement verification, our user research involved a combination of ICU ward observations and semi-structured staff interviews. To conduct observational work and recruit staff to interviews, the lead researcher (AT) partnered with a Junior Clinical Fellow (AR) for this project, who she accompanied and 'shadowed' on four of their regular work shifts. This included one short (8 hours, 8am-4pm) and three long shifts (13 hours) split into two day-shifts (8am-9pm) and one night-shift (8pm-9am). The observation days spanned two different ICU settings (the main hospital ICU and a private unit) to observe existing catheter placement procedures, ward dynamics, and data documentation practices. Through our presence on the wards as well as additional hospital contacts of the broader research team, we were able to recruit 15 hospital staff to our interviews. These were either held in a vacant hospital room during observations days, or conducted as remote calls using Microsoft Teams software. 

\subsection{Ethics}
The research study was carefully reviewed and monitored for compliance and privacy regulations; and approved by the NHS Health Research Authority (REC reference: 22/HRA/4824). Informed consent was sought by all participants in writing prior to the study. 

\subsection{Participants} 
Our interview participants reflect a range of professions and included predominately junior and more senior clinicians and nurses from ICU care, as well as two Stroke clinicians and three radiographers, who were reporting X-rays across the hospital more generally. The sample also presented a mix of staff who had been in their professional role for 0-2 years (\textit{n} = 8), with the remaining describing 2-5 (\textit{n} = 2), 5-10 (\textit{n} = 2) or more than 10 (\textit{n} = 3) years of experience. Gender was balanced across the study cohort with 8 self-reporting as \textit{female} (7 as \textit{male}). Each participant has been given a unique identification number to protect their anonymity, reported as: D1-D8 for doctors in ICU or Stroke care, N1-N4 for ICU nurses, and RR1-RR3 for reporting radiographers. See Table \ref{Table: Participants} for participant details. 

\begin{table*}

  \caption{Participants professional role, care setting, amount of years they had been working in this specific role, and their gender; alongside details on the location and duration of the interview. }
  \label{Table: Participants}
 
  \small
  \begin{tabular}{lllllll}
    \toprule
    Code & Professional Role & Setting & Time in Role & Gender & Location & Duration\\
    \midrule
    D1 & Junior Doctor & Intensive Care (ICU) & 0-2 years & male & Teams Call & 60 mins\\
    D2 & Junior Doctor & Intensive Care (ICU) & 0-2 years & male & Teams Call & 60 mins\\
    D3 & Junior Doctor & Intensive Care (ICU) & 0-2 years & female & On-Ward & 60 mins\\
    D4 & Junior Doctor & Intensive Care (ICU) & 0-2 years & male & Teams Call & 60 mins\\
    D5 & Registrar & Intensive Care (ICU) & 0-2 years & female & On-Ward & 45 mins\\
    D6 & Consultant & Intensive Care (ICU) & 2-5 years & female & On-Ward & 20 mins\\
    N1 & Staff Nurse & Intensive Care (ICU) & >10 years & female & On-Ward & 40 mins\\
    N2 & Senior Staff Nurse & Intensive Care (ICU) & >10 years & male & Teams Call & 35 mins\\
    N3 & Senior Staff Nurse & Intensive Care (ICU) & 5-10 years & female & On-Ward & 45 mins\\
    N4 & Charge Nurse & Intensive Care (ICU) & 0-2 years & female & On-Ward & 35 mins\\
    D7 & Registrar & Stroke Care & 0-2 years & male & Teams Call & 65 mins\\
    D8 & Consultant & Stroke Care & >10 years & male & Teams Call & 60 mins\\
    RR1 & Consultant Reporting Radiographer & Imaging Department & 2-5 years & male & Teams Call & 65 mins\\
    RR2 & A\&E Superintendent \& Reporting Radiographer & Imaging Department & 5-10 years & male & Teams Call & 60 mins\\
    RR3 & Consultant Reporting Radiographer & Imaging Department & 0-2 years & female & Teams Call & 55 mins\\   
    \bottomrule
  \end{tabular}
\end{table*}

\subsection{Interview Procedure}
Our staff interviews were aimed at 1 hour in duration with some variation depending on staff availability, especially for on-ward meetings, which we needed to accommodate more flexibly (\textit{MD} = 60 mins, \textit{M} = 51 mins, \textit{min} = 20 mins, \textit{max} = 65 mins). The interviews were semi-structured and involved three main parts that sought to better understand: (1) NGT end-to-end workflow and data documentation practices; (2) challenges with existing processes; and (3) AI opportunities to assist NGT verification via CXRs. Following the capture of basic demographic information, the first part asked participants to describe the current NGT placement and CXR verification process step-by-step; about their specific role and responsibilities within that workflow; and how various steps are being documented (where and by whom). The second part of the interview explored the difficulties or problems that participants faced at any stage of this process. We asked them about their reasons for
why NGTs might be misplaced and not detected; and how they could prevent or minimize delays in identifying misplaced NGTs. For staff who checked CXRs to confirm NGT placement, we also inquired about situations of doubt or ambiguity in examining the image to understand what kind of help could assist their assessments. Whenever relevant, we asked staff to give specific examples to better illustrate their experiences.

Lastly, we tried to spark participants imagination about prospective AI by asking "what-if" questions to probe more specifically into different ways in which AI could provide insights to the NGT CXR verification process. Specifically, we explored five AI proposals (Table \ref{Table: AI proposals}) that would span three main categories of intended AI uses in clinical practice: decision support; prioritization; and task automation~\cite{zajkac2023clinician}. The proposals were further informed by radiology AI research describing functionality to: detect critical image findings~\cite{seah2021effect}; prioritize radiologist reading lists~\cite{baltruschat2021smart}; produce image segmentations for lines and tubes~\cite{frid2019endotracheal, wei2023catheter}; and generate radiology reports from medical images~\cite{Hyland2023maira, yu2023evaluating}. Translating these to the use context of NGTs, we thus proposed for AI to either (1) provide an (early) alert for a misplaced tube that could be presented to clinicians, or (2) prioritize the image in the radiology reporter worklist. We also imagined AI could (3) act as a cross-checker of image assessments made by an image reporter to flag up any potential errors (e.g., if it identifies discrepancies between human text report and what it detects from the image). The AI may also (4) offer a visual overlay or segmentations that highlight the tube line and tip on the CXR image, or it can trace key anatomical landmarks to assist image interpretation. Lastly, the AI may (5) auto-generate a (preliminary) report of the NGT CXR to speed up clinician review or radiology sign-off on. To engage in deeper conversation about the prospective use of such AI, we asked: How would you use this AI information if it was available to you? In what scenarios do you think having this AI functionality would be beneficial? What are the advantages and disadvantages? We also inquired about any concerns that the participants might have regarding any of these suggestions.

We acknowledge that different AI model types and development approaches can lead to varying capabilities and limitations that can profoundly influence their success in the world; and are mindful that each proposals come with different AI requirements (e.g., classification of NGT placement is likely an easier task to realize technically than auto-report generation). However, at this early problem investigation and formulation stage, we deliberately chose to remain more agnostic to any specific model configuration or performance. We only assume that the AI models can handle multi-modal data such as radiology images, clinical texts, or other EHR/meta data (e.g., ~\cite{Hyland2023maira, tu2023towards}. The AI proposals serve as examples to explore different system goals and factors that may enhance or reduce clinical utility.

\begin{table*}
  \caption{Overview of five different proposals for how AI could come into assisting NGT CXR verification practices.}
  \label{Table: AI proposals}
  \begin{tabular}{l}
    \toprule
   \textbf{AI proposals for assisted NGT CXR verification}  \\
    \midrule
    (1) (Early) NGT misplacement detection to alert clinicians to speed up correction\\
    (2) (Early) NGT misplacement detection to prioritize for radiology reporting \\
    (3) AI cross-checker for detecting errors in human assessment of the NGT CXR\\
    (4) Segmentation overlay of NG tube + tip/ key anatomical landmarks to guide visual assessment \\
    (5) Auto-generate (preliminary) report of NGT for clinician review or faster radiology sign-off \\
    \bottomrule
  \end{tabular}
\end{table*}

\subsection{Data analysis}
The lead researcher kept a study diary, taking notes of her observations as well as capturing room set-ups, relevant machinery, and data artefacts with photography. Any hospital or person identifiable information were either cropped from the image or blocked out as a black bar on any imagery to protect patient and staff anonymity. 

To better understand how the NGT verification process was captured in data, the lead researcher also spent time during her ward visits to review (under supervision) selected ICU patient records to better understand the timings, data location, and format for key events such as: feeding tube insertion; CXR ordering; CXR capturing and upload to EPIC; NGT verification confirmation; NGT CXR reporting; any NGT corrections; and NGT-based feeding documentations. This review served to better understand what types of data is commonly generated as part of the overall NGT position qualification process; to check for consistency in entries; and assess how existing data entries could be leveraged as potential outcome metrics (e.g., reduction in delays to feeding) for any prospective AI pilot/ deployment study. 

Observational accounts and data review results were used to meaningfully contextualize the write-up of the interview findings. For this, all interview conversations were audio-recorded. The lead researcher fully anonymized the recordings and transcribed them with assistance of Microsoft Teams in-built transcription software. Final transcripts were checked and edited for correctness and then subjected to Thematic Analysis~\cite{braun2012thematic}. This involved an intensive familiarization with and coding of the data, and their iterative organization and development into high-level themes. This analysis was guided by our main research aims of understanding opportunities and challenges for developing clinically meaningful AI for NGT placement verification via CXRs within real-world hospital workflows. 

\section{Background: Study Context \&  Process of Feeding Tube Qualification via CXR}

\subsection{Hospital context}
The research was conducted at the University College London Hospitals NHS Foundation Trust that serves as a major teaching hospital, presents a world-wide centre for medical research, and is well-known for its provision of first-class acute and specialist services. Our investigation focused particularly on AI integration within intensive care units (ICU). NGTs are amongst the most commonly used catheters in critically ill ICU patients; aiming to support the sickest, more dependent patients in the hospital in getting treatment faster or more safely. The hospitals’ ICU has 48 inpatient beds (including individual patient rooms for those with infectious diseases). It encompasses a main ICU, which is divided into two separate parts: North and South (23 beds). Three additional ICUs are located in the same building and other hospital sites, which includes a private ICU ward for specialist oncology patients (10 beds). During the day, each ICU, or its parts, has a care team comprising of: a consultant, a registrar and a team of junior doctors. Overnight, one consultant is on-call at home and a registrar looks after the unit (or its two parts); alongside a team of junior doctors. Each patient also has a bedside nurse 24/7 dedicated to them. Despite our studies primary focus on ICU care, we also included the review of existing practices in the hospitals’ Stroke department as another inpatient area that frequently places NGTs, to explore how gathered insights and proposed avenues for AI may translate and generalize across different care settings. Here, the approximately 40-inpatient Stroke cohort (excluding patients with Stroke on ICU) is split between a hyper acute unit, and longer-stay patients, who are not in the hyper acute stage of their care. Nursing staff ratios differ depending on patient acuity and range between one nurse to four patients, to one nurse to eight patients. To be able to better understand the clinical applicability of AI capabilities for NGT misplacement detection as well as to clarify broader opportunities and challenges for integrating AI into radiology workflows, the next section provides an overview of the end-to-end ICU workflow of NGT placement verification via Chest X-ray.

\subsection{End-to-end ICU Workflow of Nasogastric Feeding Tube (NGT) Verification via Chest X-ray (CXR)}

\begin{figure}[p]
\includegraphics[width=20cm, angle=90]{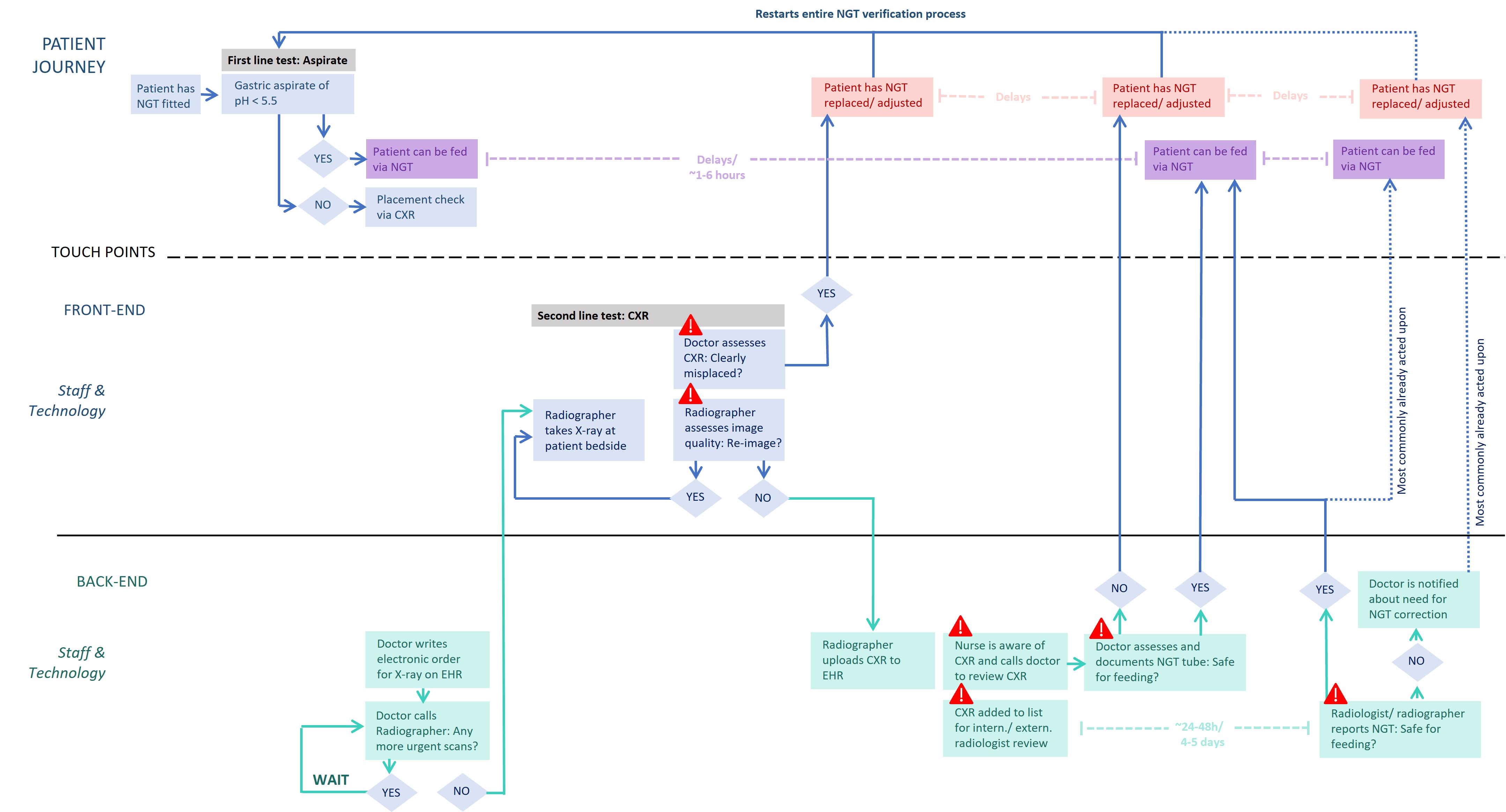}
\caption{Overview of ICU end-to-end workflow of CXR based NGT  verification from initial insertion through to decisions to use NGT for feeding or correct it, including temporal delays. The red exclamation mark symbol illustrates areas of potential integration of AI assisted critical NGT misplacement detection at different time-points and for different user groups. Note that this workflow differs from Stroke care where clinicians are not present at CXR capture, nor often review the image themselves, and rely on official radiology reporting. In ICU, NGTs are most commonly already acted upon by the time of the official radiology report.}
    \label{fig:workflow}
\centering
\end{figure}

%This section gives an overview of the ICU end-to-end workflow of qualifying the placement of a nasogastric feeding tube (NGT) via chest radiography; from its initial insertion through to decisions to use for feeding or to correct the NGT. 
On ICU, unless the patient arrives with an NGT already inserted (e.g., by theatre surgeons), the decision to place such a tube is a complex, multi-faceted process that considers the patients: (i) nutritional needs; (ii) ability to swallow that is needed for oral feeding; (iii) acuity (e.g., if a patient is very unstable, feeding is often not a priority), and (iv) other risk and comfort factors (e.g., insertion could cause delirium). Given other, often more acute patient conditions, patient feeding may be less 'urgent' and NGT insertion tasks or placement checks be delayed within a targeted 24-48 hour period from devising an NGT feeding plan to the patient being fed (see Figure \ref{fig:workflow} for a workflow overview). 

NGT placement is typically conducted by an ICU nurse. Here, any difficulties to insert the tube and physical patient response (e.g., oxygen saturation drop) can already surface a potential misplacement. To verify correct placement, there are various methods: This includes the rare use of a laryngoscope during placement, which directly visualizes the tube location. More commonly, and in keeping with the hospitals detailed NGT insertion and position verification policy, the NGT verification process involves as first line test: the \textit{obtaining of gastric aspirate}. This means a syringe is used to acquire gastric content from the stomach via the placed NG tube. This content is then checked for its pH value; a pH value of 5.5 or less confirms the NGTs location in the stomach, deeming it as safe to use (cf., ~\cite{ PolicyandGuidelineCommittee_2023}). At this hospital, staff described however how aspirate-based pH test however were often only successful in 8-10\% of ICU cases, since many patients receive proton pump inhibitors (cf.~\cite{temblett2013ph}) like pantoprazole -- a type of antacid medication that contributes to generally higher pH values. As a result, for most NGT placement checks, and especially upon any newly placed NGT, Chest X-rays are requested as a more definite second line verification test. 

CXRs for NGT confirmation are requested by ICU doctors via the hospital’s EPIC\footnote{https://www.epic.com/} electronic patient record (EHR) system, followed by the doctor using a pager to contact the on-call radiographer, awaiting call back to discuss the case and schedule the CXR based on urgency triage with other imaging requests. Since the majority of ICU inpatients are too unstable to be transported to the imaging department, CXRs are mostly carried out via a mobile X-ray machine that is brought to the patients bedside by a radiographer, who captures the CXR as an anterior-posterior\footnote{In this imaging position and view, the patient has a metal plate placed behind their back and the radiation beam traverses through their front chest to the back of their body} (AP) image. The captured image immediately appears on a preview monitor of the X-ray machine. The path and tip of the NGT may show on the machine display and is typically reviewed by the patients’ doctor. At this time, the doctor and radiographer ensure image quality of the performed CXR is acceptable, and if not, this might initiate repeating the CXR acquisition. Whilst not intended as formal assessment, ICU doctors and radiographers may already identify misplaced NGTs at this time. However, no verbal assessment of the image is accepted as grounds for the actual feeding decision. Instead, decisions that the NGT is safe-to-feed need to be formally documented in writing within EPIC once the image is uploaded to the hospitals’ Picture Archiving Communications System: PACS~\cite{fennell2021pacs}. Rather than being automatic, this image upload however requires the radiographer to physically connect the mobile X-ray machine to the hospital computer system, a process that can be delayed if there’s a sequence of scheduled X-rays that the radiographer needs to perform. The ICU NGT CXR capture process differs from Stroke care, where patients are rarely imaged at bedside and are instead taken by a Porter to and from the imaging department (ID) – a process that requires additional resources and organization. It also means that in this alternative workflow, the patients’ clinical care team (e.g., Stroke physician) is not present during image capture. Yet, image upload is instant.

\begin{figure}[h]
\includegraphics[width=13cm]{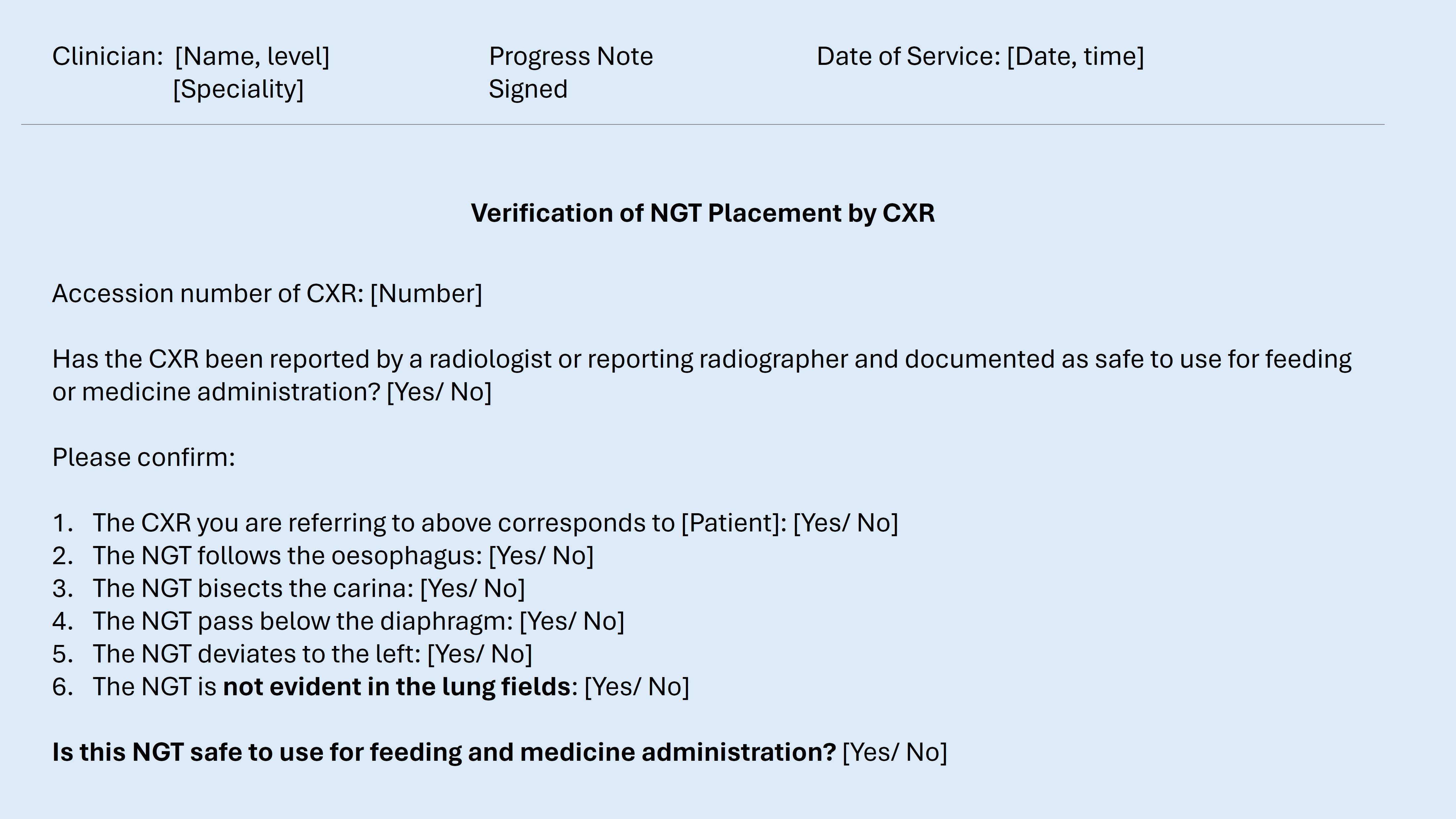}
\caption{Schematic representation of the `.NGT' template used to verify NGT placement outside an official radiology report.}
    \label{fig:template}
\centering
\end{figure}

Once the patient CXR is uploaded to PACS, it is added to a queue of images awaiting official radiology reporting by a radiologist or reporting radiographer. The large number of CXRs obtained each day, especially in intensive and emergency care, however means that image interpretation can be substantially delayed~\cite{yi2020computer}. As a result, it is common practice for ICU doctors and more senior (Stroke) doctors to check the CXR and verify the NGT’s correct positioning and suitability for use~\cite{subramanian2019automated} prior to the radiology report being issued~\cite{tang2021clip}. To this end, ICU bedside nurses, who frequently review all incoming patient results, tend to notify doctors that a CXR image is available on the system and ask for its official documentation. Doctors then log into PACS to review the image, using zoom and image contrast enhancement tooling to aid their image assessment. To document NGT placement, ICU and senior Stroke doctors utilize a smart text template, called `.NGT', that provides them with a set of binary questions that they complete in the absence of an official radiology report, and that gives clinical permission to commence patient feeding if deemed as safe. The .NGT template questions (Figure \ref{fig:template}) serve as visual check points to ensure correct placement of the NGT, which in most normal patient cases is defined as a tube that follows down the oesophagus, bisects the carina, passes below the diaphragm and then deviates to the left such that the tube tip is located in the stomach. 

\begin{figure}[h]
\includegraphics[width=13cm]{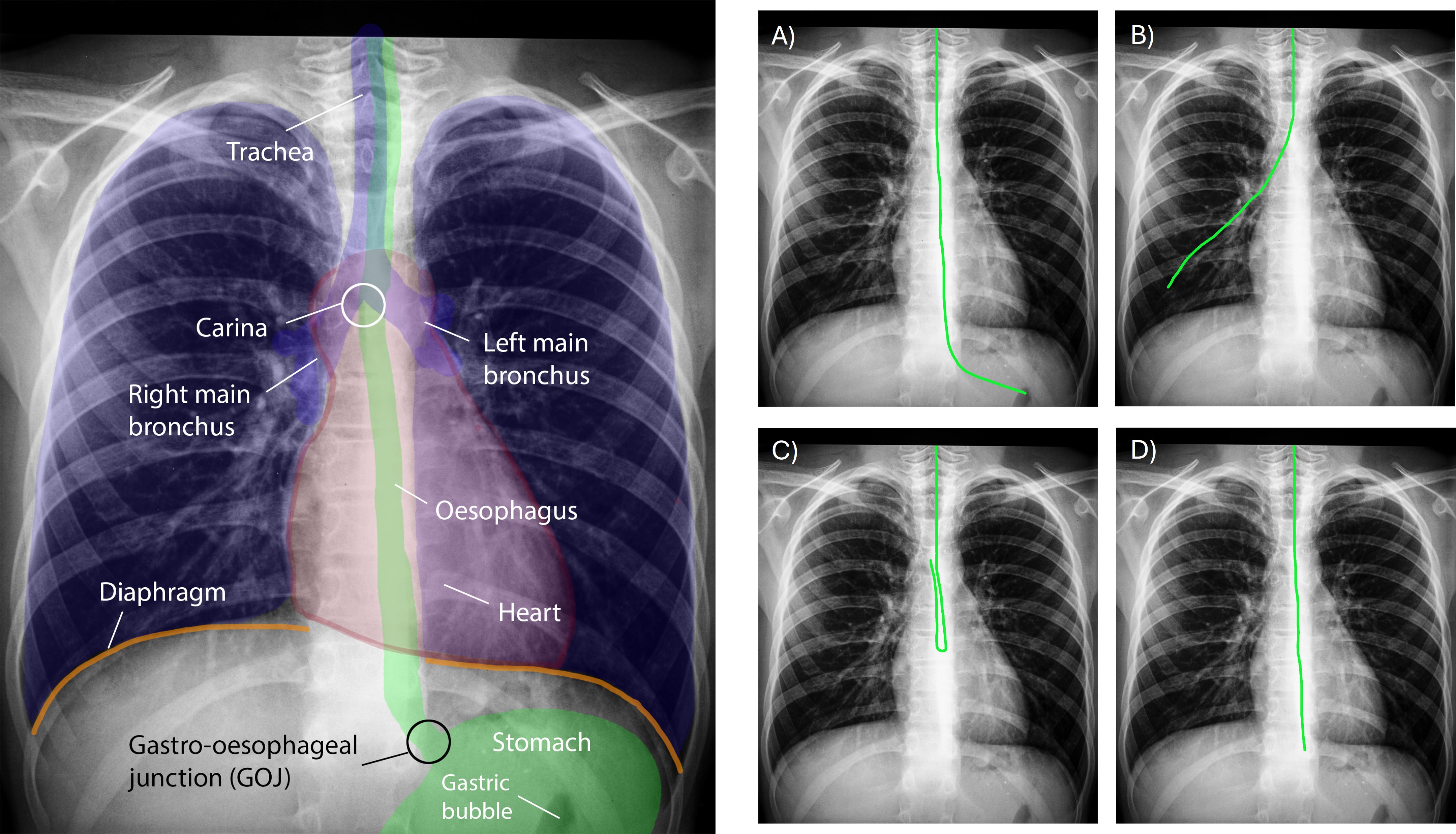}
\caption{Left: Simplified illustration of a patient’s key anatomical structures of relevance to NGT placement assessment (the heart is drawn for clarity only). Right: different NGT placements: A) correct placement of NGT with tip projecting in the stomach; B) critical misplacement of the NGT into the patient's right lung (via the trachea and the right main bronchus); C) misplaced NGT which is coiled in patient’s mid oesophagus; and D) misplaced NGT with tip in distal oesophagus requiring further advancement to reach the stomach.}
    \label{fig:NGTcases}
\centering
\end{figure}

Importantly, an NGT should not be placed into the patient's lungs. Feeding a patient through an NGT misplaced into the lungs would be a severe incident that can have critical implications – including patient death – and is classified by the UK National Health Services (NHS) as a Never Event\footnote{Never events are “serious incidents that are entirely preventable because guidance or safety recommendations providing strong systemic protective barriers are available at a national level, and should have been implemented by all healthcare providers”: https://www.england.nhs.uk/patient-safety/revised-never-events-policy-and-framework/”}. Less critical, yet sub-optimally placed NGTs may not extend far enough into the stomach; need withdrawing; or may be kinked or coiled along the path – thereby inhibiting proper use and intended functionality. See Figure \ref{fig:NGTcases} for illustrations of normal, sub-optimal and critically (mis)placed NGTs. Any intervention to correct or replace the NGT then restarts the entire NGT verification process (Figure \ref{fig:workflow}). 

%Once the doctor has reviewed and documented the NGT CXR they tend to inform the nurse, usually in person, about its safety for feeding or need for correction.  
Once an NGT is correctly placed, ICU nurses can start patient feeding following either a standard protocol or a more bespoke plan that ICU dieticians devise for the patient based on their weight and caloric needs. While the hospital in general aims to have inpatient X-rays reported within a 12-hour period for acutely unwell or emergency care patients (ideally under 4 hours during normal work-hours)\footnote{https://www.england.nhs.uk/long-read/diagnostic-imaging-reporting-turnaround-times/}, actual turn-around times are much longer (e.g., more than 7 days), which eliminates its practical utility and relevance in clinical decision making. Reporting times differ in other care settings, like Stroke, where junior doctors are not permitted to access CXRs and are reliant on the official report by a radiographer/ radiologist or a more senior clinician. Still, delays are common as turn-around times vary depending on reporter availability and can be significantly delayed for non-acute CXRs and requests made outside main work hours (e.g., at night, weekends). 

Where CXRs are assessed by non-radiology staff, and when working in stressful care environments with staff and resources stretched to capacity, image interpretation is prone to human errors~\cite{elaanba2021automatic}. This suggests potential utility of leveraging AI image analysis capabilities, for example, to \textit{alert clinicians to critical placement} or \textit{have NGT CXRs prioritized for urgent radiology reporting}; enabling earlier detection of misplaced NGTs to speed-up their correction and prevent any additional complications~\cite{tang2021clip}. AI may also assist in \textit{detecting human errors }in image assessment, \textit{provide visual guidance} to human interpretation, or help \textit{automate NGT assessment} as captured in our AI proposals (Table \ref{Table: AI proposals}).   

\section{Findings}
Table \ref{Table: Learnings} summarizes our study findings that center on answering two main questions: How to best leverage AI to assist the NGT CXR verification process in a clinically meaningful way (e.g., improve workflow effectiveness, or patient safety)? How does understanding the use context and available data provide important insights to effectively guide AI  development and prospective (system) study design? Responding to these, we structured our findings into two parts: 

\begin{table*}

  \caption{Summary of findings and implications for developing and evaluating clinically relevant healthcare AI.}
  \label{Table: Learnings}
 
  %\small 
  \fontsize{7}{8}\selectfont
  \begin{tabular}{p{0.15\textwidth} p{.8\textwidth}}
    \toprule
    \multicolumn{2}{l}{\textbf{Clinically Meaningful AI Applications}}\\
    \midrule
    
   Scope  &  
   \begin{itemize}
       \item Given broader design goals (increase radiologist effectiveness, improve patient safety), remain open to identifying alternative, potentially more important opportunities for AI applications (e.g., beyond a focus on image analysis). \textcolor{gray}{\textit{Ask: What are the most important problems to focus on? How could AI be best placed to assist those?}}
       \item Consider how clinical value and risks of different AI proposals depend on their realization in practice. \textcolor{gray}{\textit{Ask: How well will AI need to perform to realize its potential? How well can we assist AI verification/ error detection? What are the implications of undetected AI errors? For what types of use cases would an imperfectly performing AI still provide utility?}}
       \item Clarify medical accountability of AI assisted clinical practice. \textcolor{gray}{\textit{Ask: Is the purpose of the AI to “augment” or “automate” (new/ existing) human practices? What is the added value proposition for use cases that require AI use “only with human oversight”? What are medical-legal and broader organisational requirements? }}
   \end{itemize} \\
     \midrule
   Target users  &  
   \begin{itemize}
       \item When clarifying target users for AI, consider opportunities \& implications of (re)defining AI-assisted role responsibilities including training requirements, overall workload burden, and clinical responsibilities. \textcolor{gray}{\textit{Ask: How generalizable may an AI application be across care professions and care settings (e.g., different wards, hospital environments)? Are there any (future) changes anticipated for certain roles/ professions?}}
   \end{itemize} \\
     \midrule
    Workflow integration/ design choices  &  
    \begin{itemize}
       \item Be mindful of frictions introduced by new (safety) practices or substantial changes to existing workflows that AI might introduce (e.g., requirements of time, training, changes to routine). \textcolor{gray}{\textit{Ask: How does the new, AI approach sit alongside, or presents an improvement to, other current (safeguarding) practices?  What would be the simplest implementation of AI that causes least disruption to existing work, but would still be insightful?}}
       \item Map out where within an end-to-end workflow AI could be incorporated and its implications to more systematically guide choices (see Table \ref{Table: Map} as an example). \textcolor{gray}{\textit{Ask: Where are areas of opportunity to assist the workflow and what would be the benefits and risks of different AI implementations to direct and indirect stakeholders?}}
       \item Balance requirements for seamless workflow integration with needs for AI transparency and verification. \textcolor{gray}{\textit{Ask: How to  introduce the right types of friction (e.g., through careful alerts/ notification frequency based on good-enough AI performance) to ensure greater safety balanced with avoiding overburdening or delays? What types of AI applications or their design allow for easy, fast verification and correction by humans?}}
   \end{itemize} \\
     \midrule
       \multicolumn{2}{l}{\textbf{Data Constraints \& Opportunities}}\\
     \midrule

    Data labelling  &  
   \begin{itemize}
       \item Investigate how data generation and documentations may differ from previous data(sets). \textcolor{gray}{\textit{Ask: What additional data processing stages may need to be performed (e.g., to anonymize data, extract valid labels)? What are existing, standardized data capturing processes that could be leveraged (e.g., to assist data labelling)?}}
       \item Consider differentiating between data items that are clear (ground truth) vs. more ambiguous to label.  \textcolor{gray}{\textit{Ask: What are more factual labels that could be introduced that avoid interpretation variances?}}
   \end{itemize} \\
     \midrule
     
    Model training/ evaluation  &  
   \begin{itemize}
       \item Clarify relevant patient factors (e.g., BMI, consciousness) and patient edge cases (e.g., hiatus hernia, situs inversus, gastro/ oesophageal surgery), and other confounders (e.g., image quality) for the AI task.  \textcolor{gray}{\textit{Ask: What other relevant (meta/ EHR) data should be included in model development or evaluation?}}
       \item Map out different AI error types with regards to the use case to test model performance against. \textcolor{gray}{\textit{Ask: What error types might occur \& how can these be (automatically) detected in data evaluations?} What are the implications of different error types for users assessment of the image (or other clinical data)?}
   \end{itemize} \\
   
    Evaluating AI application success &
       \begin{itemize}
       \item Consider real-world data generation practices and their limitations to clarify clinically relevant success criteria. \textcolor{gray}{\textit{Ask: What can be evidenced reliably with the data that is available? What additional data sources need including?}}
       \item Define realistic measures for evidencing AI success by considering the broader use context. \textcolor{gray}{\textit{Ask: What moderator variables  (e.g.,  hospital dynamics related to resourcing, shift patterns, emergencies) may implicate staffs ability to timely act upon AI insights and may need including in evaluation study protocols?}}
   \end{itemize} \\
      \midrule
       \multicolumn{2}{l}{\textbf{Broader Implementation/ AI Adoption Considerations}}\\
     \midrule

     AI expectations, training \& organisational framing &
       \begin{itemize}
       \item Help clinical stakeholders develop realistic expectations of AI. \textcolor{gray}{\textit{Ask: What are current expectations of AI capabilities and how well do they map to technical performances? For what use cases is AI more and less likely to assist with?}}
       \item Devise plans for staff training \& onboarding (see ~\cite{cai2019hello} for guidance), and clarify AI (pilot) deployment or broader roll-out strategy with hospital/ Trust leadership.
   \end{itemize} \\
   \bottomrule
  \end{tabular}
\end{table*}

\section*{PART A: Identifying Clinically Meaningful AI Applications for NGT CXR Verification}
Our investigation into meaningful AI NGT applications begins with (1) insights into existing workflow challenges, safeguarding mechanisms and their limitations. We then detail how (2) perceived clinical utility of proposed AI functionality is bound-up by a complex interplay and decisions surrounding: the target user group; desired AI outcome prioritization; workflow integration and AI design choices; as well as broader AI acceptance and adoption challenges. Based on these learnings, we conclude this section with (3) an example of how identified factors of application type, workflow stage, user group, design choice and technology performance requirements can be mapped to clarify and trade-off the potential clinical impact and feasibility of different AI proposals. 

\subsection{Existing Workflow Challenges, Safeguarding Mechanisms \& their Limitations}
This section first outlines broader workflow challenges surrounding the end-to-end NGT CXR verification process that expand the scope for AI opportunities as well as give important context for data work. We then specifically describe the prevalence of errors, their reasons, and existing safeguarding mechanisms in, NGT image assessment and subsequent safe-to-feed decisions, which presents the main focus of our AI investigation.

\subsubsection{Delays to Timely NGT Placement Verification as Most Prevalent Workflow Problem} 

\hfill \break
When asked about key challenges or pain points in the NGT verification process, rather than describing difficulties in CXR image interpretation, staff most prominently mentioned the length of `time' that it takes, and the various types of `delays' that span the entire process from initial NGT placement, through to the CXR being scheduled, captured, uploaded, reviewed and documented, and the patient eventually being fed. We learned that timeliness of NGT placement verification is bound up by a complex interplay of: (i) staff resourcing and emergency-led care requirements; (ii) communication inefficiencies; and (iii) broader ward dynamics. 

\textbf{\textit{Resource Constraints \& Emergency-led Care that Implicate Staff Ability to Act upon NGT Tasks:}} With hospitals operating on “bare minimum of staff for everything, particularly at night”, staff availability (and level of expertise) required to execute NGT-related tasks is especially limited at weekend and out of main working hours. This affects ICU staff, who can be busy  or caught up in a crash call\footnote{In a crash call on ICU, all doctors are required to be at hand for a deteriorating patient who experiences a cardiac arrest and needs resuscitation}. Similarly, image capturing or reporting staff can be in high-demand. For example, one radiographer alone may need to cover both CT and X-ray imaging, which then often leaves NGT checks last due to a need to prioritize more urgent patient cases, as well as its balancing with needs of other busy hospital departments: 

\begin{quote}
\small
    “(…) it's all [based] on this understanding of emergency need. And so X-rays to check NG positioning sit awkwardly in the way that most hospitals plan their overnight resources, or all of their acute flow, because they're not quite acute, and yet at the same time, they need to be done soon.” (D8, Stroke)
\end{quote}

Low staff resources and emergency-centric care mean that X-rays may not get timely reported unless acute. Consequential delays, often by several days, however mean that their clinical relevance, especially for assessing a misplaced NGT, or otherwise to guide clinical examination is lost. As a result, ICU staff tend to “not wait” (D5), “rely on” (D4), nor “check” (D2, D3) the radiology report for NGTs:  

\begin{quote}
\small
 “Yeah, I guess the main thing is if you're not immediate reporting them, are they really of any use to anyone? That’s the thing, isn't it? Unless you're getting a definite result right there, is there any point of reporting something that's three days old and that they're already using? I don't know. I guess that's the one of the main challenges.” (RR3, ID)   
\end{quote}

\textbf{\textit{Communication Inefficiencies \& Constantly Chasing-up Tasks:}} 
Along the NGT process, staff described human checks (e.g., keeping taps on patient records to identify if a report has been issued) or in-person verbal exchanges whereby nurses `chase-up' doctors about the need to order an X-ray or document it, alongside electronic messages or system notifications. In their accounts, they describe communication inefficiencies such as risks of missing digital messages and notifications on EPIC in good time; incomplete handover information across staff teams that delay actions; and radiographers forgetting to carry the bleep or to call back (D1); as well as doctors missing call-backs, forgetting to order, or being late to review the CXRs. All this suggests the need for a more reliable communication system and better assistance with administrative burden:

\begin{quote}
\small
    “The frustrating parts of being a doctor is how much time you spend doing kind of admin tasks such as being on the phone all the time, waiting for bleeps back from people, or waiting to bleep people, waiting for reports of things to come, waiting for emails. All of these types of things which AI might make our lives quicker and easier in, and free us up to do the decision-making part of the job or the assessment part of the job would be really beneficial.” (D1, ICU)
\end{quote}

\textbf{\textit{Ward dynamics of Task Offloading based on Staff Confidence, Flow \& Shift Handover Times: }}
Timeliness of NGT task completion and their prioritization are further moderated by (i) staff’s ability to safely execute a task (e.g., their competence, time availability to place the NGT or do checks). More implicitly, staff described their awareness of strain on under-staffed services, which they responded to by offloading, for example tasks from night shifts to the day team. Describing considerations of heightened risks in conducting certain, less urgent procedures at night due to increased tiredness and awareness of a pending shift hand-over, a junior doctor stated:

\begin{quote}
\small
    “So, so this all happened overnight and I remember being very busy nightshift where I think there were only a few hours left of the night shift. And we were like: actually, this is probably safer left to the morning team. Just don’t use it, don't touch the NG at the moment (…)” (D3, ICU). 
\end{quote}

Staffs considerations of shift patterns in decision to prioritize, hold or post-pone tasks is further evident in an account whereby an NGT is placed on ICU at night or early morning, but does not require urgent confirmation. In such cases, the doctor may suggest to the radiographer on-call that image capture could wait for the day team, being mindful that requesting an image scan close to shift hand-over (usually 8am or 8pm) can mean for a radiographer having to stay at work longer:  

\begin{quote}
\small
    “But I also know that, if they [the radiographers] hand over at 7:30 AM to their colleagues and I'm bleeping them at 7:15 and they go, it's kind of day team come. If I say no, we need it now, then they're not going home until late. So then, human nature, you're kind of like, yeah, it's fine.” (D1, ICU)
\end{quote}

Task prioritization is further balanced with staffs desires and needs to protect the “flow” of clinicians and nurses in completing tasks. The below quote illustrates how a senior ICU staff nurses (N1) is conscious to not want to interrupt doctors too frequently about various tasks. Instead, she keeps a “rolling list” of things that are less urgent that she then informs them about upon their next exchange: 

\begin{quote}
\small
    “I mean, the doctors are in and out anyway, so I tend to have a little rolling list of things I want to tell them. So I don't forever be interrupting them in their flow as well. If some thing's can wait till the evening ward round, I just kind of wait and then just say this needs doing, you know, and then we can get it all sorted in one go. Otherwise, it's a bit fragmented for them. They're kind of jumping between patients all the time, and it's quite…They lose their flow of thought as well. I don't think all nurses think that way, though I don't think more junior ones would think I need to get this sort, I'm gonna talk to the doctor and the are forever talking to the doctor about things, but I can kind of prioritise what they need to know straight away, yeah.” (N1, ICU)
\end{quote}

All of this draws attention to complex human factors and work rhythms that determine the prioritization and timeliness of NGT task completion and data documentation. These broader workflow dynamics have two main implications. Firstly, suggested delays and inefficiencies in existing NGT verification broaden the scope for how AI could assist current radiology practices above and beyond image interpretation (e.g., focused on optimizing communication/ triage management processes). Secondly, interdependencies caused by variations in staff resourcing; emergencies; or shift pattern present key variables that may hinder staff’s ability to act upon specific AI insights (e.g., an AI alert to a misplaced NGT) and therefore require considerations in any evaluative studies (see further Section 5.7).

\subsubsection{Prevalence of, Reasons for Errors and Existing Safeguards in NGT CXR Image Interpretation \& Safe-to-Feed Decisions}
\hfill \break
Outside process delays, and given the serve implications of missing and potentially using an NGT that has been critically misplaced into the patients' lung, we next report staffs’ accounts on: the prevalence of NGT misplacement; reasons for errors in NGT assessment; and current hospital risk mitigation strategies. These serve to better understand where AI could add value; and may sit alongside, or replace existing safeguarding practices. 

While staff confirmed that feeding tube placement occurs frequently on ICU and Stroke wards, NGTs were described as rarely misplaced. On occasion, NGTs were found to be sub-optimally placed, whereby the tube is either coiled, needs advancing or pulling back; most commonly as a consequence of difficulty placing the tube or the patient dislodging a tube that is already in place. Its critical misplacement into the lungs, however, was described as very rare, with an estimate of 2 cases within a 6-week period on ICU. Almost all of these tend to be spotted prior to any feeding, meaning that so called `never events' rarely ever occur. In fact, the majority of participants reported to have only ever encountered such a case once or twice – most often with no direct involvement (e.g., reported by a colleague, occurrence on the ward/ department).  
Asked to speculate about reasons for why a critically or sub-optimally placed tube was not spotted in time was attributed most often to (1) human error whereby either the wrong image was reviewed or the image was not assessed correctly; as well as (2) additional (technical) challenges pertaining to poor image quality/ tube visibility and also specific patient factors and edge cases. Next, we detail these challenges, their contributing factors, and describe existing safety procedures and their limitations for ensuring safe image assessment. 

\hfill \break
\textbf{Human Errors in Image Assessment. }Errors in image review were commonly attributed to the person reviewing the CXR being unable to think clearly due to a busy, understaffed shift, or tiredness during the night. Errors are also bound up with lower image reader expertise and confidence and lack of proper adherence to hospital policy and safeguarding processes (e.g., documentation protocols). Whilst more senior clinicians (e.g., registrars, consultants) generally described checks of NGT position as one of the easiest, most straightforward CXR assessment tasks, more uncertainty was expressed for more junior, less-experienced staff (D2, D3, D5, RR2), who would commonly ask for peer review. A reporting radiographer reflects:

\begin{quote}
\small
    “(…) when I was first starting, even though I felt like I was sure it was in the right place, I was never confident it was always in the right place, so I'd always ask for someone else to check, but I think over time, once you've done enough of them and you’ve unfortunately seen ones that go into the lung, you can discern between ones that are in the right place and aren’t in the right place a bit more.” (RR2, ID)
\end{quote}

To support confident image review practices, our analysis surfaced three main risk mitigations: (i) cultivation of a mindset of caution to prevent patient harm and frequent engagements in human peer review; (ii) mandatory, standardized reporting on EPIC via a template that enforces key visual checkpoints; and (iii) requirements for nurses to check the safe positioning of the NGT on the patients’ nose and conduct regular aspiration checks at the beginning of their shift, and at 4-hourly intervals.

\textbf{\textit{Mindset of Caution \& Peer Review:}} Entrusted with their patients’ care, staff for example described being cautious and vigilant, and to ask for a second opinion or conduct extra checks if they had any doubt about the NGT’s position to avoid potential mistakes. An ICU nurse reflects on this error-preventing mindset:

\begin{quote}
\small
    “So we just need to give them [the patients] topmost care, so I think each and everything, we don't need to go drastically. We need to take second opinion or some others opinion. Because if I have any doubt then I just need to stop there. I just need to escalate and I just need to take a second opinion on something like that. And we have a lot of ways to check. Because if anything wrong happened, at that time, we can't do anything. So there is a word that prevention is better than cure. So we just need to prevent everything.” (N2, ICU)
\end{quote}

Outside of specific reasons for uncertainty, it was generally also regarded as “good practice” (D4), especially for junior doctors, to consult ideally a senior colleague to review their assessment. In fact, some of the junior doctors stated they engaged in peer review “every time” they needed to document an NGT (e.g., “I just want to be 100\% on this type of thing” (D1)). Yet, the ability to connect with a more senior staff as peer review can be more difficult at night or on a busy ward, which may potentially force a more junior, tired doctor to make a decision without additional human safeguards: 

\begin{quote}
\small
   "There could be emergencies in the unit, so the registrar and the rest of the team are busy with that, but your patient needs that NG tube confirmed. So that's when a potential never event could happen because the pressure of you need[ing] to confirm it but you don't have a person to second ask. So you might confirm it just for confirming sake.” (D2, ICU)
\end{quote}

\textbf{\textit{Mandatory, Standardized Documentation: }}Hospital policy on NGT verification further requires to have written documentation that the NGT is safe for feeding on EPIC, which serves to ensures a doctor reviews the image, and possibly as a legal trace. Outside the official radiology report, this documentation requires clinicians to use an NGT specific smart text reporting template (see Figure \ref{fig:template}).  designed to mandate responses to six image assessment checkpoints to reduce chances of interpretation errors. Despite the existence of the NGT template, and its predominant use over free text reporting (which also exists), a recent audit conducted by one of the junior doctors (D2) revealed that only ~83\% of NGT templates had all questions completed. Describing the implications of negligence in properly following the review protocol, the doctor remembers a case where a senior doctor did not trace the tube path to check if it bisected the carina and went down the oesophagus; while the tube tip looked like it was below the diaphragm in the area of the stomach, the clinician missed that the tube had pierced side-ways through the lungs instead (D2). Three clinicians further remarked on the irritating design of the final visual assessment question through its enforced negation, which risks greater confusion and more likely the giving of a false answer.

Although mandatory requirements of written NGT documentation and template adherence can increase safety, we found that such safeguards also invite friction and compete with desires to also speed-up clinical decisions and patient feeding. In this regard, one ICU registrar describes her perception of documentation and training requirements for NGT confirmation to be “a lot of over-doing” and as not necessary for clear cases:

\begin{quote}
\small
    “In my personal opinion only, it's a lot of over-doing. I've worked in other places and it wasn't done like that. We would put the NG feed in, we would use what is known as the whoosh\footnote{The whoosh test is a method whereby air is rapidly injected down an NGT while auscultating over the epigastrium. Listening to the resulting sound, a gurgling indicates air entering the stomach, whilst its absence suggests the tip of the NGT is elsewhere (e.g., lung, oesophagus)~\cite{dawson2007nasogastric}}, which is basically blowing air inside and listening to see if it, or feeling that it's in the stomach, if it's there and if you can aspirate anything out of it, then it's fine. We wouldn't document it. It's like a routine procedure that we do, we wouldn’t document it. (...) I'm saying if there's a doubt and we would also order a Chest X-ray and we will always document what we saw in the X-ray. But we don't have like a special… I never used a special entry template. I never had to go through like a test to make sure I can do that. It’s taken for granted that if you are a doctor that you are supposed to be able to look at a Chest X-ray, you don’t need to be tested for it.” (D5, ICU)
\end{quote}

\textbf{\textit{Continuous NGT Position \& Aspiration Checks: }}Following NGT placement confirmation, ICU bedside nurses check the NGT daily, at every shift, and at 4-hour intervals, to look for any signs of displacement. Changes to measures of the tube length and checking if it is still securely attached to the nose can be key indicators that an NGT may have moved and be misplaced. Yet, such NGT measurements are not error-proof as a tube can, i.e., coil in the mouth or oesophagus: 

\begin{quote}
\small
    “The concern to me is that if it's sub-optimally placed, advancing and then I often ask them to re-advance it to check that it's not coiling. And that's often people will try to advance it and actually, the coiling applies in the mouth, so we've reached a sort of problematic, anatomically, down there, in that placement.  So coiling above the X-ray position. (…) You generally can see that if you put 58 centimeters in and it has just gone into the mouth. You can generally see that, but there are occasions where you advance it, and instead of advancing at that point, all that happens is you create a kink point within the mouth. I mean this is infrequent, this is but it's. So that's often why I would want to advance and check rather than just advance. So if its too short you would want to check that it has actually moved rather than… An awake patient would tell you there’s something weird in my mouth, but it's the intubated ones.” (D6, ICU)
\end{quote}

Furthermore, ICU nurses responsibilities include `4-hourly aspiration' checks of the NGT as continued confirmation of its correct placement where possible. If no aspirate could be obtained, the result will be counter-checked by a second person, and, should results not match, a third checker. If in doubt, any feed that had been started would be stopped and the patient send for X-ray. Across the ICU and Stroke setting, CXR was perceived as a more definite, reliable test for assessing feeding tube placement rather than aspirate (e.g., “Everybody would rather do an extra Chest X-ray to check absolutely that tubes in the right position” (N1)). Especially in Stroke care, aspirate checks and required 4-hourly wait periods in-between unsuccessful tests were criticized for holding up X-rays and causing overall delays. Expressing frustration about the process and how aspirates, if not obtained first time, most likely won’t be obtained second time, the Stroke registrar shared:

\begin{quote}
\small
    “And I also think, this is perhaps a bit more philosophical, but I also think that in introducing like little barriers and delays into something that is very essential, both to get right and also essential not to get wrong. It's really dumb to have delays. It’s really dumb to frustrate people with the process. In other places, the guidelines have been able where it's like if there's any uncertainty, get a chest X-ray. And I actually really don't understand quite how much of a barrier it is in this organization. I mean I can understand why it's evolved, but it wouldn't be my chosen thing. But it is what it is.” (D7, Stroke)
\end{quote}

While the above processes, alongside staff training, are sought to increase safety, engagements in peer review; requirements for written NGT documentation; and (multiple) aspiration checks prior to CXR orders were also attributed as sources of delays. Within a hospital culture that is already perceived as very safety oriented – with a clear NGT specific policy, staff training and skills tests – this surfaces friction whereby desires for patient safety actively compete with desires to also speed-up clinical decisions and patient feeding; raising questions if and how AI can meaningfully fit within, risk competing with, or needs to be thought of as better alternative to those existing safeguarding practices. 

\hfill \break
\hfill \break
\textbf{Technical Challenges in Image Assessment (Unclear NGT Visibility \& Moderating Patient Factors): }Finally, we learned that image assessment difficulties and subsequent mistakes can be rooted in technical challenges pertaining to: poor image quality or tube visibility due to insufficient image penetration; difficulties to fully trace the tube path and its tip on the CXR; or external artefacts obscuring its view. Other patient factors, such as obesity and other opacities at the lung base (e.g., consolidation) that show as white on the image – alike the NGT –  further hinder a clear view of the tube. Where the patient is (hugely) rotated, it can also: “give the impression that [the] NG tube is in a different place, so you're not getting the true position of it” (RR2). To trouble shoot those image quality and assessment difficulties, clinicians predominantly described improvements to image capture or viewing via: (i) the use of better X-ray machines, (ii) improved image capture settings or imaging modality (e.g., via CT), and (iii) review on higher-resolution monitors. 

Lastly, we learned that patients can have an unusual anatomy, for example due to lung pathology, esophageal or stomach surgery. Where their anatomy has changed (e.g., due to a gastric pull-up or a gastroesophageal resection), correct NGT placement verification can be more difficult and usually requires a more individual assessment, oftentimes involving consultations with other clinicians (e.g., gastrologists) and reviews of the patients (image) history to get necessary context information to aid NGT assessment. In patients over the age of 50\footnote{Hiatus Hernia https://www.nhs.uk/conditions/hiatus-hernia/}, a condition called hiatus hernia is also very common, whereby their stomach moves up into their chest and can show above the diaphragm. Whilst deviating in their position from the standard protocol (Figure \ref{fig:template}), the NGT can still be correctly placed into the stomach and be safe for feeding. Similarly, in very rare cases, a patients’ stomach can be to the right, rather than left, caused by a congenital abnormality called situs inversus\footnote{Situs Inversus https://www.ncbi.nlm.nih.gov/pmc/articles/PMC8901252/} whereby the person’s major visceral organs are in a reverse position. As detailed in Section 5.6, this understanding of image quality constraints, patient edge cases, and relevance of important patient history information present critical insights to data work (e.g., pointing to potential data biases) in this space. 

\subsection{Perceived Clinical Utility of AI: A Complex Interplay of Multiple, Interwoven Goals and Constraints}
This section details how perceived clinical utility of proposed AI functionality is bound up with (i) the target user group and (ii) AI outcomes that are being prioritized; (iii) workflow integration and AI design choices; as well as (iv) broader AI acceptance and adoption challenges. 

\subsubsection{Determining the Right Target User Group for the AI}
Reviewing the ICU NGT workflow, the question surfaced, who is commonly reporting NGT CXRs and would benefit most from specific AI capabilities to assist NGT assessment in their work? Based on our understanding of the end-to-end workflow, this potentially involves: ICU clinicians and nurses (or senior Stroke staff); imaging and reporting radiographers; and radiologist/ radiology registrars.

\textbf{Lower Involvement of Radiologists in NGT CXR Workflow:} Contrary to our initial assumptions that placed radiologists as the most likely user group (e.g., to prioritize clinical findings in their reading list), we learned that interactions with radiologists or their reports rarely featured within the ICU NGT CXR verification workflow. ICU clinicians and reporting radiographers would only reach out and call a radiologist if they had a very specific, urgent clinical question, which rarely occurs for any NGT-related queries except for very complex or edge cases (e.g., gastric surgery or very obese patient). This differs in Stroke care, where junior doctors are not allowed to assess the X-ray to make a safety-critical feeding decision. Instead, they have to refer to a more senior staff, most commonly the registrar, to confirm and document the NGT. In the absence of senior colleagues, Stroke doctors -- in contrast to ICU doctors -- therefore rely on the official radiology report. This official reporting is either provided by few (\textasciitilde8-9) reporting radiographers, or by radiology registrars or radiologists, who report the NGT CXR as part: of acute reporting; radiology registrars training; and to increase reporting capacity (there are \textasciitilde120 consultant radiologists employed at the hospital). In general, there was a sense that X-ray reporting, especially for NGTs “distracts the radiologist, who's reporting, from other things” (D8), such as more complex imaging (e.g., CT, MRI) or diagnostically more challenging tasks that require their specific expertise; suggesting CXR reviews for NGTs could be offloaded to other professions:

\begin{quote}
\small
    “[About prioritizing critically misplaced feeding tubes in the reading order of radiologists] “I think that's not a bad idea. Only tricky thing about that is how many could they get and a lot of the time when it's not, I think obviously speaking from my experience, but I think a lot of the time when it is misplaced, it's quite clearly misplaced. So I'm thinking is it worth taking a radiologist time to check or can a clinician, trained looking at NG tubes, like for any of the SHOs\footnote{SHO stands for Senior House Officers and represents an umbrella term for multiple doctor grades in the UK.} on critical care, just have a quick glance at it before escalating it to a radiologists.” (D4, ICU)
\end{quote}

\textbf{Offloading to Senior/ ICU Clinicians or Upskilling ICU Nurses:}
As alluded to in the above quote, doctors with permission to verify NGTs may present a suitable target user group of an AI alert to critical misplacement to speed up its correction especially where official radiology reporting is otherwise delayed. While ICU doctors are all specifically trained and usually the first people to respond in assessing CXR images to verify NGT, this is not common practice across all hospital care settings (e.g., it differs in Stoke), and may not easily generalize to other hospitals or Trusts. 

We also explored if ICU nurses, as the ones who place the NGT, could potential become the recipient of an AI alert. ICU nurses described to have a good understanding of any difficulties surrounding the NGT placement, and often had already build-up some understanding of recognizing NGTs on CXRs.  While this suggests that (ICU) nurses could be upskilled to take on NGT CXR reviews, they are currently not authorized to access radiology images (and cannot access PACS), nor would they have the clinical expertise to potentially detect other medical conditions of the patient that may show on the X-ray and fall into the diagnostic remit of doctors; thereby blurring diagnostic boundaries and medical-legal responsibilities. 

\textbf{Radiographers’ Role in Early Intervention \& for Generalization Across Care Settings:}
As a user group, we distinguish the roles of imaging radiographers, who operate imaging equipment to capture X-rays (and other medical images); from reporting radiographers, who interpret and report on those captured images. It was common in the reports of different staff that imaging radiographers often spot NGT misplacements straight way, at point of acquisition (see workflow diagram, Figure \ref{fig:workflow}), when it is clinically most relevant as it can be act upon immediately – whereas a reporting clinician or reporting radiographer may not see the image until days later. Imaging radiographer can also alert nurses to the need for immediate review or re-insertion of a misplaced tube: 

\begin{quote}
\small
    “Well, obviously, I can't think of a specific example, but I can tell you exactly … sometimes it's caught by the radiographers, the radiographers are very good at this and they can spot this straight away and then they would contact the clinician straight away to inform it, because that's why, actually probably the most important thing is that it's caught straight away, because by the time we've got to us as reporting radiographers. As a reporting clinician, it could be 2 days down the line. Patient may have been fed by that point. So the crucial point, the crucial point, is at that point of acquisition.” (RR2, ID)
\end{quote}

When considering that feeding tubes can be placed anywhere in hospital and by many different staff, upskilling all those professions to NGT CXR assessment was described as unlikely to scale. Instead, imaging radiographers were suggested as the “only constant” (RR1), involved in NGT CXRs at day and night, and across any ward and hospital in the world; suggesting their upskilling to this task: 

\begin{quote}
\small
    “Also we're looking at radiographer-led as well. So we can look at doing some training for the radiographers and if they're on the ward, at night time or something like that and then they will assess it and make their judgment themselves which is not quite off the ground yet, but we're looking at it.” (RR3, ID)
\end{quote}
 
Currently, imaging radiographers without postgraduate education however are currently not authorized to provide a radiology report. Nonetheless, it is expected of their professional competence at the point that they appraise the image - should they identify critical findings on an examination -  to take direct and timely action by alerting the referrer\footnote{https://www.hcpc-uk.org/standards/standards-of-proficiency/radiographers/}. Furthermore, one of the consultant radiographers described an initiative that plans to standardize NGT reporting and build competencies for all imaging radiographers in England in the near future: 

\begin{quote}
\small
    “(…) And that intervention is aimed at standardizing training and education with uniform competency assessment for all radiographers who take chest X-rays where a nasogastric tube maybe present. So the diagnostic radiographers you would have met in ICU. The ambition is to train everyone of those, in England, to be able to identify a nasogastric tube on a chest X-ray and make a safe to feed, not safe to feed decision basis. (...) So the Royal College of Radiologists and the College of Radiographers facilitated by HEB [Health Education England] have commissioned a piece of work that is active at the moment to do that stepped process.” (RR1, ID)
\end{quote}

In contrast, CXR-based NGT checks feature high in the workload of reporting radiographers, who reported that approx. 10\% of their inpatient reports include requests to confirm NGT placement. Despite this prevalence, reporting radiographers tend to report predominantly within main-working hours, and are currently part of the main pathway for clinicians to reach out to, instead of radiologists, for any urgent, NGT CXR reviews. Consequently, as indicated in the end-to-end workflow (Figure \ref{fig:workflow}), reporting radiographers described to come too late into the image review process, stating that any misplaced NGT would already been identified by the ward clinicians. In few instances where radiographers were first to spot a misplaced tube, it was because clinicians hadn't yet looked at the images themselves.

Considering the importance of timeliness in CXR review, it is worth exploring the potential of alerting these reporters as early as possible to a critical or sub-optimal NGT. Reporting radiographers tend to report more narrowly – for example only chest and skeletal X-rays – compared to radiologists, who also report on other modalities (e.g., MRI, CTs) and subspecialities (e.g., brain, abdomen). This narrower reporting focus was suggested to be more suited to prioritizing critical findings in their reporting worklist given the smaller range of possible acute findings that could surface and otherwise compete with each other for urgency: 

\begin{quote}
\small
    “If you are a consultant radiologist working on the acute desk and you are reporting a MRI spine for cord compression. You are reporting CT chest for pulmonary embolism. You are reporting a nasogastric tube for placement confirmation, and you are reporting a CT abdomen to perforation. How many prioritization algorithms do you run? And what trumps what in a worklist? So is your bleed on a CT brain more or less urgent then your misplaced nasogastric tube, which is more or less important than your pneumothorax, which is more or less important… Where one, I'm [a] simple creature. I can report a handful of things, and therefore I'm easy to prioritize, because I do this few things.” (RR1, ID)
\end{quote}

This section illustrated prospective benefits as well as feasibility constraints of different AI target user groups. While an imaging radiographer could be supported in identifying a critically misplaced NGT as early as possible on the X-ray machine, access to vendor software to integrate AI functionality is likely more complicated; nor do current hospital policies and reporting protocols (in writing) yet permit their involvement in this verification and reporting task. As such, proposals to early alert the patients’ clinical care team of a likely misplacement, who can either act upon it themselves or mobilize a radiology review; as well as prioritization in CXR reporter work lists appear more aligned with existing care pathways. In Section 5.3 below, we continue to map out the implications of such choices on staff practices and for patient risk/benefit.  

\subsubsection{Balancing Across Goals of: Patient Safety, Assessment Confidence and Process Efficiency}

\hfill \break
Next we detail how participants perceived the potential utility of our five different AI proposals that describe desires for: improving patient safety; staff confidence in image assessment; and overall process efficiency; and we outline some of the tensions that can exist between those goals.  

\hfill \break
\textbf{Patient Safety as the Most Important Requirement. }When asked what type of AI application participants would find most beneficial; they recognized value across the various AI proposals, with some even expressing a preference for a combination of different AI capabilities to provide more effective safeguarding support in the NGT verification process. In particular, staff emphasized \textit{patient benefit and safety }as the most prominent driver for clinical care. As a result, AI functionality that (i) assists the detection of sub-optimal or critically placed NGTs – either as an alert directed at (ICU) clinicians or for image review prioritization (especially in Stroke care) – or that (ii) flags up potential reporting errors was ascribed most clinical relevance to speed up immediate problem detection. 

\textbf{\textit{Early Detection of NGT Misplacement (for ICU Clinicians or in Prioritizing Reporting Lists):}} Having an AI automatically alert or flag-up an abnormal NGT early in the verification process was described as “definitely helpful” (D7) as it (i) enables \textit{immediate action}: ICU doctors would “go and look at that image straightaway” (D4). An early misplacement alert would also (ii) invite extra “caution” (D2) and closer inspection (e.g., via peer review), which was regarded to improve patient safety by \textit{reducing risks of human or contextual factors} that can cause for a misplaced tube to be missed and the patient being misfed. All of this also\textit{ reduces overall delays to patient feeding}: 

\begin{quote}
\small
    “I think it could only be a good thing, right? Because sometimes patients are waiting days before they get fed. So if this is highlighted straight away and it's cutting down a risk of misfeeding, I think it would be great if you, even if it had to wait 5-10 minutes for somebody to put in the official report. I think it could only be a benefit really.” (RR3, ID)
\end{quote}

In Stroke care, where junior clinicians cannot verify CXR NGTs themselves but rely on the official radiology report, the proposal to re-prioritize the reading list of image reporters to speed up assessment of critical NGTs was also regarded as “extremely helpful” (D7). Reporting radiographers too described prioritizing critical examinations that needed their attention as one of the most beneficial and “appropriate” uses of AI (RR1), suggesting that amongst a list of 200 patients: “(…) if there was something that could hopefully highlight the ones that needed more urgent attention, that would be great.” (RR3). Alike early alerts to ICU clinicians, sooner reporting due to AI prioritization of critical NGTs was proposed to \textit{reduce delays in taking corrective actions}, and in \textit{subsequent feeding or medicine administration}. It was also considered to \textit{reduce risks of inappropriate self-reporting} due to formal reporting delays that otherwise could present a safety concern: 

\begin{quote}
\small
    “In practice, reasonable wards would not be feeding until they've got the report, and really the problem is delays. And in practice, whoever is reading the report, in practice good systems wouldn't take matters into their own hand and attempt to report the scan themselves, even if not competent. But I completely understand if you have a system immediately for flagging that the tube is in the wrong place, you are minimizing the chance of either feeding happening, assuming it's alright, delays to feeding and delays to medicine, and the last which would be inappropriate self-reporting because of delays to formal report and therefore wrong conclusion of safety.  All of those sound straightforward potential benefits of the technology.” (D8, Stroke)
\end{quote}

Despite its potential, an open question remains to what extend an early NGT misplacement alert could indeed speed-up clinical review or report generation – and consequentially correction times. On ICU, nurses tend to chase doctors to review the image as soon as is feasible, as they too have an urgency to work through their tasks; whilst doctors, attending to a busy ward, may not be able to review the image any earlier than they currently do. Similarly, in Stroke care, even if a critically misplaced NGT was brought higher up in the reading list of a radiographer, there may still be a long reporting gap, if a review was requested out-of-reporting-hours.

\textbf{\textit{AI for Error Checking:}} Similar to critical placement detection, having the AI flag a potential reporting error prior to the report or clinical note submission was described to invite caution and a closer look, and to serve as a useful “little backup” (RR3) and “extra layer of security” (D4) to \textit{prevent human errors} of misreading the image or accidentally clicking the wrong answer option on the .NGT template caused by rushing, distractions, or inexperience:

\begin{quote}
\small
    “(…) I think for us doctors the most important thing is patient safety as opposed to efficiency and trying to get things done fast and trying to identify things quicker, and workflows, and all these kind of things. I think for us it’s safety and the only one that really safeguards, as a safeguard, is that one option [error-checking], because it makes you take a step back and think again.” (D4, ICU)
\end{quote}

Benefits to have AI serve as a second checker of their X-ray assessment was especially valued by ICU doctors with lesser experience in assessing NGT placement, who may feel pressured to decide or may not have access or time to consult a (senior) colleague. Here, having the AI take on the role of a second checker is suggested to reduce overall NGT confirmation time as it: “would probably cut down the time that you might check with a senior colleague when you didn't have the confidence to know” (D3).

While alerting to a human error in image assessment could catch mistakes that otherwise can have significant implications on patient safety, there is uncertainty about the prevalence of actual reporting errors for NGTs. If at all detected, reporting errors are not commonly captured outside of any official critical incident reporting and documentation practices, leaving this as an open challenge for data work. 

\hfill \break
\textbf{Assisting Image Assessment Confidence as `Nice to Have'.} Having an AI produce a visual overlay on the CXR image, where its “literally highlighting where it thinks the tube is” (D3) including its path and tip, and key relevant landmarks (e.g., the carina bifurcation point), were described as helpful for \textit{improving image assessment confidence and speed} – especially in more junior clinicians; and for simplifying the completion of the .NGT template. 

\begin{quote}
\small
    “It would make the workflow much easier. Because, some of the difficult times, like, especially if it's junior colleagues, they would be delaying the documentation because they are not complete 100 percentage sure. So if AI can provide a certain amount of assurity to it and they can also visualize OK, this bisects the carina, and it's below the diaphragm then it definitely reduces the time in which it is confirmed.” (D2, ICU)
\end{quote}

Despite these proposed advantages in assisting reporter confidence and speedy reporting; there was general agreement across hospital staff that feeding tubes are commonly “quite easy” to identify (D3, D4); especially since they present the only tube that extends that far down into the body, and when compared to other, more complex neck and PICC lines. Similarly, key landmarks like the stomach, which tends to show as a gas bubble on the CXR, or the patients’ diaphragm (Figure \ref{fig:NGTcases}), are regarded as simple anatomy for a doctor to locate; all of which questioned the extend of the usefulness of an AI overlay, outside more complex patient cases: The Stroke registrar draws a parallel to other AI software: 
\begin{quote}
\small
    “Yeah, I mean, I personally I think that [the AI overlay] would be really good. Number one, I'd find it interesting, I guess another question about whether I'd find it like helpful, but I would find it interesting. I would definitely, if the image was there, I would definitely look at it. To draw an analogy with the CT scans we do in Stroke, I'm always intrigued by what the brainomix\footnote{https://www.brainomix.com/}, I don't know if you've ever seen it? I always scroll through. Basically you have the scan that we look through and then there's another like basically an extra study which is overlaid on the scan, its the AI's interpretation of the scan. And I do look through that, cause I do think it is interesting and once in a while, what it flags as an abnormality, I would have a second look at.” (D7, Stroke)
\end{quote}

Consequently, when weighing-up our various AI proposals, staff ascribed more clinical utility towards proposals that more directly and immediately aided the detection or critical findings or human errors, as opposed to boosting staff confidence or reporting speed: 

\begin{quote}
\small
    “If all that's gonna happen is you're going to make me as a reporter 1\% better. No one’s perfect, right? I'm not perfect. AI is not perfect. No one’s perfect. We accept that people get things wrong. (…) If you are just making experts a very, very, very tiny bit better, that's good, but that's not where the money is. The money is identifying or prioritising those examinations, the critical finding, who need your attention first. So that's useful for me, because I just report Chest X-rays and Skeletal X-rays. So if you identify critical findings, of either of those two things that needs to be the next thing that I report.” (RR1, ID)
\end{quote}
\hfill \break
\textbf{Increasing Reporting Capacity (or Speed), but not at the Expense of Safety:} While the above proposals of AI assistance in critical findings prioritization and AI error checking serve as safeguards to clinical assessment and to increase patient safety, they do not increase image review capacity or reduce image backlogs that contribute to official reporting delays. Here, having an AI auto-generate a (preliminary) NGT report or have it pre-fill the .NGT template could help. Yet, such proposals received mixed responses. In terms of clinical utility, participant feedback varied depending on how auto-report functionality was implemented to either: (i) \textit{assist} or (ii) \textit{fully replace} existing NGT image review practices.

\textbf{\textit{Weighing-off Benefits and Risks of AI-Assistance in NGT Reporting: }}Where AI image interpretation and reporting automation functionality is intended to assist clinicians review practices (e.g., ICU doctor, reporting radiographers) there was doubt about it providing much utility in terms of time-savings or to guide clinical decision making since clinicians “would still look at the x-ray [them]self” (D7) and spend time to check, verify, and if necessary, correct the AI-generated report. Furthermore, staff would still consult with a more senior colleague if there was uncertainty about their assessment of the image – despite AI input. Consequently, there was agreement that, as long as it remains the responsibility of the clinician or radiologist/ radiographer to sign off on the report, that they would do their “own checks” even if, ultimately, they state at the bottom to “agree with AI interpretation” (D7):

\begin{quote}
\small
    “ (…) let's say the algorithm said it's in the right place like how much confidence would I have to agree with that algorithm, or would I still going to have a look at the X-ray myself? And then, if I'm looking at the x-ray myself then how much help is the algorithm, because essentially if I'm putting my license on the line for this NG placement, then how much do I trust the algorithm to get it right? And I think that's kind of the issue. (…) personally for me, I would still always look, even if there was some sort of system in place to say that it is in the right place.” (D4, ICU)
\end{quote}

Simultaneously, if NGT auto-report functionality existed, doctors expressed concerns they might become “lazy” (D2, D3) and either might not open the X-ray at all, if the AI for instance directly auto-populated the clinical note (D2); or might not “always always always check everything on the protocol” (D3). 

\begin{quote}
\small
    “So the only problem then is there's a chance that they might not open the X-ray, we are lazy people. Maybe you need to at least open the X-ray to get it auto-populated into the thing, because if it auto populates into the note directly then we might not check the X-ray. It is a possibility, especially on busy nights and stuff like that. There's a chance that if AI does that automatically, so at least I think.” (D2, ICU)
\end{quote}

Furthermore, even if humans check and always have the “final say” on assessment (RR2), risks remain where the AI errs, but clinicians, especially more junior ones (D7, RR3), may “not [be] confident in overruling” what has been flagged by the AI (RR1), or be “brave” enough to go with their “own convictions” (RR2). 

\begin{quote}
\small
    “I think I broadly like it. If someone had asked me to look at that, the chest X-ray, and an AI report was already there. I would read the AI report. I would definitely read; I wouldn't just ignore it. I would. And I would still look at the image itself, look at the image with my own eyes. I guess the question comes in is, if there is a more junior or less confident colleague who looks at the AI report, is less confident about disagreeing with the AI report. If they look at the picture themselves, where does the responsibility lie?” (D7, Stroke)
\end{quote}

\textbf{\textit{Accounting for Medical-legal Responsibility of Full Automation in NGT Reporting:}} Where AI use is envisioned as providing good enough performance to automate NGT assessment or reporting, this was regarded to usefully offload work and safe ICU clinicians time: 

\begin{quote}
\small
“I can tell you that if the machine was doing it [the NGT assessment], then it would offload us from doing it and then nurses will just read the machine thing and start using the NG. So the machine would write that the NG is safe to use and it will spare me 50 minutes of my day.” (D5, ICU). 
\end{quote}

However, for this proposal to come to its fruition, it would  require for the AI to also take over "full responsibility” for the medical assessment. One of the junior ICU doctors expressed:

\begin{quote}
\small
    “(…)  if we're giving the sole responsibility of a patient to a computer program, then fine, but if we're not, then I don't really see where, apart from being a helping hand, I can't see it being that useful, if that makes sense, and that's kind of my take on it.” (D4, ICU)
\end{quote}

Simultaneously, AI automation use cases like automated image assessment and reporting also face regulatory and legal challenges that limit its practical realization:

\begin{quote}
\small
    “At the moment that is illegal, it's not compliant with radiation legislation in the UK. So anything will only ever be of clinical decision support or a triage tool. Triage and decision support doesn't create capacity, it just reallocates existing capacity because you still have to report everything.” (RR1, ID).
\end{quote}

\subsubsection{Workflow Integration \& AI Interaction Design Considerations}

\hfill \break
This section illustrates how the utility of different NGT AI proposals is bound up not only with its integration within existing workflows, and also human-AI interaction configurations and design choices that surface tensions between desires for seamless, unobstrusive AI integration and needs to check and verify the AI. 

\textbf{Enabling Seamless AI Workflow Integration. }In our conversations about AI integration within current practice, staff described the need for efficiency and seamlessness in workflow integration to not impede reporting speed or flow. For instance, in the context of reading list prioritization, reporting radiographers described the desire that high priority cases, like a misplaced NGT, would be flushed “straight up to the top of our list, so it comes up as our next patient” (RR3) to not “interrupt my workflow” (RR1) and instead to auto-load the next case, once review is finished and signed:

\begin{quote}
\small
    “My personal preference, again this is not science, is that having a report prioritisation to top of work list is useful and so it doesn't interrupt my workflow. So I'm reporting a case and finished my report. It then auto-prioritises into the reading worklist that when I hit sign, and then it auto loads the next case for me, that that case is the next one that it loads. So seamless prioritisation in the worklist is really important. I don't wanna have to keep going in and out of things.” (RR1, ID)
\end{quote}

Reflecting on the placement of an AI alert to a critical findings, their proposals varied from notifications upon opening the patient page within EPIC; through to its integration as an AI tab within the image viewer PACS; and annotations directly to the radiology image to bring extra caution to image review practices. By way of example, one of the registrars drew on historical radiographer practices whereby they would put a “red dot” on a physical X-ray image to indicate that a scan was obviously abnormal; which made doctors second check. He described how in some hospitals this practice still existed digitally, through a white text box stating "Red dot” or an “asterix” that is put on the image, indicating: “(…) This study contains a critical result. I find that helpful, so it's not intrusive, but it's unavoidable if you're looking at the scan.” (D7, Stroke).
While direct image annotations can bring key AI outputs to the forefront, in the context of AI producing a visual overlay of the NG tube, reporting radiographers also remarked on the importance for AI outputs to not interfere with their visual search strategy through the image that might prohibit the finding of other injuries/ pathologies. Instead, they expressed  preferences for the AI to show as a toggle or second image within a separate AI image tab – as is commonly seen with other AI software (D7, RR2).

Furthermore, design considerations such as (i) the timing or ordering of AI results within human workflows; or (ii) the level of clinical AI interpretation of the image are bound up with risk and concerns about AI over-reliance, and human skill acquisition: 

\textbf{Designing Human-Led AI Interactions.} For instance, reflecting on how AI functionality, like an error checker would become integrated, participants expressed a preference for doctors having to make their assessment first, before receiving an AI response to not interfere with their own clinical reading, nor with the formation of key medical skills and competencies in CXR image assessment:  

\begin{quote}
\small
    “So if we say it's a very good system, but it's not perfect. And it wasn't generating huge amounts of work, as in, there weren't masses of discrepancy in general between what it [the AI] was finding out and what you were finding, then having an AI safeguard would be helpful, yes! I think if you've done it first. ..and I think what would also be useful in that scenario is that as a clinician you wouldn’t deskill, you would still be forced to look at the image. You would still have to make a decision. You wouldn't rely on technology to make that decision for you. So you still are competent in assessment, which I think is a skill.” (D3, ICU)
\end{quote}

As mentioned above, any requirements of clinicians to make their own assessment at first, or having to double-check the AI generated (report) output, this however can limit proposed AI efficiency and utility gains.

\textbf{Focusing on Lower-Level AI Interpretation.} In the context  of how best to articulate a critical AI finding, most clinical value was ascribed to an understanding whether the tube is unsafe for feeding. In this regard, one of the radiographers expressed a preference to communicate the AI output as a binary finding (Is it safe to feed: Yes/ No) without any context why the AI made that decision such that he still needs to critically appraise the X-ray. This is sought to reduce automation bias and over-reliance on the AI (RR1).

\begin{quote}
\small
    “The most important question that AI would be able to answer is whether it's unsafe for feeding. So for example, if it's not bisecting the carina, the AI would be able to pick it up and it could potentially prevent a never event from happening.”  (D2, ICU)
\end{quote}

Others preferred for the AI to only flag-up and alert towards an abnormally placed NGT rather than for it to interpret the image (e.g., by stating about the NGT that “it’s safe” to use (D7)) to reduce risks of any negative implications in instances where the AI was incorrect. Given that the role of image reporters is to interpret the image and to “only say what they see (…) [not] decide what should be done with the patient” (D5), this suggest the role of AI to be better placed and lower-risk in assisting image assessment; leaving any more interpretative medical conclusions – whether feeding is safe or not – to the clinical team. 

\textbf{Emphasizing Human Skills Acquisition. } Many participants saw potential of using AI, especially the visual overlay, to assist in NGT training or skill development scenarios. For example, they proposed for AI to: (i) show the “optimum position for the tip” (RR3) on the image; (ii) learn where tube tip needs to be (how far advanced) to provide guidance to junior doctors; or (iii) link to demonstrations of the correct way to textually describe the NGT. Simultaneously, they also raised concerns about how AI use might complicate skills acquisition. Reflecting on the question of how novices of today become experts of tomorrow, staff emphasized the importance of developing appropriate internal feedback loops that ensure clinicians won't repeat certain mistakes. They also cautioned not to train people to report only with AI, which risks AI over-reliance. To better illustrate how AI use could risk de-skilling staff, one ICU registrar drew a parallel to ECG auto-impression, which may lead staff to not take the time to think, and can remove opportunity for developing important image interpretation skills:

\begin{quote}
\small
    “I can tell you on the ECG today, the machine is calculating all kinds of stuff and coming up with diagnostics, like writing ‘patient has this’. I never read it. Because it's very often wrong, and because it makes you not think. The same like when you download an article. When you make ChatGPT write your assignment, then you didn't learn anything. Maybe I’m old-fashioned. (...) if all my life I looked at what the machine was interpreting I would never know how to interpret an ECG. (...)” (D5, ICU)
\end{quote}

\subsubsection{AI Acceptance more Broadly: Setting and Managing Appropriate AI Expectations}

\hfill \break
Lastly, clinical utility of our AI proposals for assisting NGT placement verification is bound-up with AI performance requirements as well as overall AI acceptance and adoption by medical professionals and health organisations. 

\textbf{Need for High AI Performance: Surpassing Human Capabilities for AI Utility.} Across AI proposals, chances of the AI being incorrect or to miss important instances were described as the “main risk” (D3) and “worry” (D4); even if an AI system was “really, really good” (D7). Especially for more ambitious AI concepts such as auto-generation (or auto-completion) of the NGT report (template), participants expressed “doubts” (N2) and a lack of trust in AI’s capabilities, suggesting it being useful: “if we can show that it is 100 percentage accurate” (D2), which is likely an unrealistic expectation of AI to fulfil. Probing deeper into what would be considered as good enough performance for AI to take full responsibility in NGT assessment, one of the ICU clinicians suggested it needed to be “better than a human interpreting” (D4) – alike computers that beat human chess players (D5): “It may mean that we may trusting something that may fail, but humans fail. So who is more likely to fail?” (N4). As described above, in cases where technology performance at least surpassed human capability, full-automation in NGT reporting could usefully offload work and save ICU clinicians time (if medical-legally permitted). 

\textbf{Costs \& Disruptions Incurred by Imperfect AI.} Even for less risky AI uses, such as an AI error checker application, clinicians emphasized the need to achieve good enough performance to not overly constrain human processes in cases where it is incorrect. Technically, cases where the AI may flag an otherwise unnoticed human error, should only occur very rarely. Thus, flagging-up disagreements in cases where clinicians were competent to make the assessment, or the AI was incorrect, can cause additional overhead and reporting delays (e.g., requiring additional peer review); and may mean AI alerts are becoming ignored going forward: 

\begin{quote}
\small
    “What I don't know is, how many times do qualified professionals say an NG tube is appropriately placed and then the AI disagrees. How often does that come up? Because it might be quite annoying if I look at 10 X-rays, and all of them, the AI disagrees with me, but it all ends up being like they're fine. In which case people just gonna click through that alert and like, never bother with it.” (D7, Stroke)
\end{quote}

\textbf{Desired AI Performance: Theoretical Ideal vs. Reality.} When discussing areas of opportunities for AI, we observed tension between clinicians suggesting its use to support instances that were difficult for humans to assess – for example, have AI “add an extra layer” of insight (D4) that could save clinicians having to call a radiologists, thereby reducing demands on their time – especially for more complex patient or poorer image quality cases – whilst simultaneously expressing doubts in AI capabilities to perform accurately and reliably.  

For example, most clinical utility for a visual AI overlay was ascribed to recognizing the NGT path in \textit{more complicated cases} where patients either: (i) have had a poor posture during image capture (e.g., ICU patients are often slumped badly, N4); (ii) have multiple tubes or lines – internally or externally – overlying their chest, complicating their differentiation; (iii) have other pathologies (e.g., big effusions) that can complicate the identification of key landmarks such as the diaphragm; or (iv) present with an unusual anatomy. There is an open question however, how well AI would perform in those more complicated, often edge cases. Given that even expert clinicians and official radiology image reporters can be unsure about the NGT placement in these instances (e.g., asking for better image quality), this may also limit their ability to verify and contest the AI output, posing further risks in cases where AI results could (more likely) be false. 

In general, our findings indicate how it can be difficult for clinicians to frame realistic expectations of AI.  For example, while one ICU clinician could see the prospect of AI making their “life easier” (D5) and described new AI capabilities (like ChatGPT) as “extraordinary” (D5) she was mindful this technology is still new, thus regarding such AI capabilities as more of a \textit{theoretical proposition rather than a practical reality}: 

\begin{quote}
\small
   “(…) and it's still just starting, so it's not perfect, I mean. Right now, AI is also making up a lot of nonsense. It's just not true. Yeah, but here we are discussing [a] theoretical thing. If it works, good, I would like to have it. If it works 50\% and the other times it’s lying completely, then I wouldn't wanna use it.” (D5, ICU).  
\end{quote}

Participants poorer performance expectations and lack of trust in AI capabilities are partially grounded in experiences with other prior technical innovations. Making reference particularly to patient electrocardiogram (ECG) readings that have an auto-generated impression of the ECG signal printed at the top of paper strip  (e.g., “This ECG shows ST elevation across the anterior leads”), ICU doctors described how they would still make up their “own opinion” of the ECG, even “recalculating” the machine outputs (D4), or not reading its impression at all as they found it many times to be “wrong or out of context” (D5).

\begin{quote}
\small
    “So I think that goes back to what I was saying about ECG, because that's essentially what the ECG machine does. It gives you a preliminary report, but despite that, every time I'd still look at the strip myself and even some of the calculations, I recalculate them by hand. Because, I mean, at the end of the day, obviously the ECG machine works more on plan recognition as opposed to obviously using big data or what, so it’s quite hard to trust exactly what’s being presented to you. (…) It does get it wrong sometimes. So you see things that are reported or preliminary reports on the ECG strip and you're going: well, that's clearly not what's going on with this strip. So I think that's where the double check is needed unfortunately.” (D4, ICU)
\end{quote}

\textbf{Broader Considerations for Managing AI Expectations \& Acceptance. }Lastly, our participants also touched on broader considerations for how AI acceptance and adoption would need to be achieved, including: (i) need for AI education for staff to develop appropriate AI literacy; (ii) requirements for effectiveness and practice integration studies to better understand the (negative) effects of AI use on people’s reporting; (iii) desires for carefully staged approaches to AI performance assessments and its continuous monitoring; and (iv) clarity on Trust policy on AI responsibility and how much certainty clinicians can give to the AI outputs:

\begin{quote}
\small
    “I think it really depends on what the Trust decides is like a level of certainty that we give to the AI report. Because I think if junior doctors were told the AI report is just a guide, but it's not sufficient, they would have a very different attitude to it, for if they were told we have faith in the AI guide. If you're not sure, speak to a senior. I think those like there's, like, very different responsibility burdens of proof, depending on how the AI is framed.” (D7, ICU)
\end{quote}

All these considerations shape appropriate expectations of AI and how it becomes adopted within clinical practice. 

\subsection{Mapping Intended AI \& its Potential Implications to Distill Clarity and Facilitate Discourse}
The above research insights paint a complex picture of the design space for AI-assisted NGT CXR support within an ICU hospital workflow. In this section, we exemplify via Table \ref{Table: Map} how those insights can be mapped out in more systematic ways to help research and development teams create more clarity across different intended AI uses;  design configurations; and their implications on direct and indirect stakeholders to better weigh-off prospective AI benefits and risks. 

For the use case of AI assisted NGT misplacement detection, our example mapping shows how seemingly the same functionality enabled by AI -- \textit{the detection of a (critical) NGT misplacement} -- could become incorporated at different workflow stages. For each of three example instances depicted in Table \ref{Table: Map}, the various implementations would involve a different target user group (imaging radiographer vs. ICU clinician vs. radiologist/ reporting radiographer), different software integrations (e.g., X-ray machine, EPIC alert, PACS viewer) and design choices (e.g., alert notification vs. worklist reordering). To help think through the potential benefits and risks of each instantiation, we chose to map desired AI performance of correctly identified (critical) misplacement events against likely errors (e.g., false classification of correct NGTs as misplaced), and reflected on its direct impact on staff (practice) and indirectly, on patient outcomes. This clarified that the use case likely offers patient benefits, with comparatively lower risks as long as AI performance is optimized to avoid the misclassification of correctly placed NGTs. 

The break-down also shows that clinical relevance and patient benefits of a misplacement alert are likely highest at the point of image acquisition and may more easily scale across care settings (e.g., beyond ICU) given the consistent role and presence of radiographers at CXR capture. However, current technical and medical legal-constraints suggest lesser practical feasibility for this instantiation. Whilst reporting radiographers and radiologists are most qualified to appropriately review CXRs and critically appraise AI outputs (to avoid risks of AI over-reliance), their often late reporting of ICU NGT CXRs suggests that ICU doctors may be the most beneficial target user group to act upon a timely alert for greater clinical utility. 

It is worth noting that our mapping example suggests a simple, binary AI misplacement classification with a restricted scope of possible errors. For other use cases, such mapping would benefit from extending. For instance, an AI-generated NGT visual overlay may surface AI errors such as: missed, false or broken path tracing; non-, or false indication of tube tip; or NGT misclassified as a different tube type. For NGT report generation, error types likely extend even further and may include i.e.: errors in distance quantification when specifying tube location; or errors in safe feeding classification. Likely those different error types can have more or less severe implications if remained undetected. For instance: a technical mistake in misclassifying a \textit{naso}gastric tube as an \textit{oro}gastric tube (a feeding tube that is inserted via the mouth, not nose) that is nonetheless correctly identified in the area of the stomach is unlikely to pose any significant risks to patient safety. This suggests the need to bring closer attention to developing and deploying more methods that help formalize and systematically assess the (likely) clinical implications of system errors above and beyond technical performance metrics.

\begin{table}

  \caption{Example of mapping a proposed AI functionality across different workflow stages; illustrating variances in user groups, design choices, potential benefits and risks of desired versus erroneous AI outputs on staff and  patients; and broader context considerations.}
  \label{Table: Map}
  
 \rotatebox{90}{
  \small
    \fontsize{6}{7}\selectfont
  \begin{tabular}{p{0.1\textwidth} p{0.05\textwidth} p{0.05\textwidth} p{0.1\textwidth} p{0.1\textwidth} p{0.25\textwidth} p{0.25\textwidth} p{0.2\textwidth}}
    \toprule
    AI Capability & Workflow Stage & User & Design Choice & AI Performance & Potential Implications on Staff/ Practices & Potential Implications on Patients & Broader considerations\\
    \midrule
    Detection of (critical) NGT misplacement & CXR preview image & Imaging radiographer & Alert on (mobile) X-ray machine & Correct critical or sub-optimal misplacement detection & 
    \textit{Benefits} 
    \begin{itemize}
        \item Earliest possible review of CXR to detect NGT misplacement (when its clinically most critical)
        \item Informing clinical team to review for sooner NGT documentation and correction
        \item Most Consistent user group across care settings (all CXRs beyond ICU)
    \end{itemize}
    &  \textit{Benefits} 
    \begin{itemize}
        \item Highest impact on reducing risk of missed NGT critical (or sub-optimal) displacement + related complications
        \item Reduced delays to patient feeding via more timely correction
    \end{itemize} 
    & \textit{Technically feasibility}: requires AI implementation within radiology equipment software 
    
   \textit{ Medical-legal constraints: }imaging radiographers often not permitted to report NGT CXRs 
    \\
     &  & &  & False alert: NG tube is correctly placed  & 
    \textit{Risks} 
    \begin{itemize}
        \item Alert fatigue
        \item Unnecessary disruption to doctor/ nursing workflows 
        \item Increase in time spent in image review or peer review causing delays 
    \end{itemize}
    &  \textit{Risks} 
    \begin{itemize}
        \item Other (potentially urgent) patient imaging / care becomes delayed 
    \end{itemize} 
    & \textit{AI performance: }avoid misclassification of correct NGTs in model optimization  
    \\
    
     \midrule
    & CXR image review & ICU doctor & Critical findings alert in EPIC & Correct critical or sub-optimal misplacement detection & 
    \textit{Benefits} 
    \begin{itemize}
        \item More immediate review of CXR image on PACS
        \item Earlier detection + action to correct critical NGT misplacement 
        \item Reduced risks of critical NGT misplacement remaining undetected (due to human or contextual factors)
    \end{itemize}
    &  \textit{Benefits} 
    \begin{itemize}
        \item Reduced risk of missed NGT critical displacement + related complications 
        \item Reduced delays to patient feeding via more timely correction
    \end{itemize} 
    &    \\

    &  & &  & False alert: NG tube is correctly placed  & 
    \textit{Risks} 
    \begin{itemize}
        \item Increase in time spent in image review or peer review causing delays 
        \item Risk of misinterpretation of tube location 
        \item Confusion about AI output/ loss in AI trust 
    \end{itemize}
    &  \textit{Risks} 
    \begin{itemize}
        \item Other (potentially urgent) patient imaging / care becomes delayed 
    \end{itemize} 
    &    \\
    
     \midrule
    & CXR reporting & Reporting radiographer/ Radiologist & CXR image with (critically) misplaced NGT is prioritized at top of PACS reading list & Correct critical or sub-optimal misplacement detection & 
    \textit{Benefits} 
    \begin{itemize}
        \item Earlier official reporting of (critically) misplaced tube that reduce risks of inappropriate self-reporting (e.g., by more junior doctors)
        \item Reduced risks of critical NGT misplacement remaining undetected (due to human or contextual factors)
    \end{itemize}
    &  \textit{Benefits} 
    \begin{itemize}
        \item If detected prior to clinical team: reduced risk of missed NGT critical displacement + related complications 
        \item Reduced delays to patient feeding via more timely correction
    \end{itemize} 
    &  \textit{Workflow: }Often (in ICU care), clinical team will have already acted upon NGT CXR prior to official report especially if reporting only happens within-hours  \\

    &  & &  & False alert: NG tube is correctly placed  & 
    \textit{Risks} 
    \begin{itemize}
        \item Competes with other critical urgent findings that can become deprioritized
    \end{itemize}
    &  \textit{Risks} 
    \begin{itemize}
        \item Other (potentially urgent) patient imaging / care becomes delayed 
    \end{itemize} 
    &  \textit{ Prioritization:} Difficulty managing different, potentially competing “critical findings” in urgency tirage  \\
    
    \bottomrule
  \end{tabular}
  }
\end{table}

\section*{PART B: Data Opportunities \& Challenges for AI Development and Evaluation}
The last section of our findings details learnings about existing NGT hospital data - it's production, availability and characteristics  - for NGT-specific AI development, and closes with a review of relevant outcome metrics and their limitations, if a prospective NGT AI application was to be deployed and studied within healthcare practice.

\subsection{Understanding NGT Data Requirements \& Data Quality/ Characteristics}
Extending on previous descriptions on how poor image quality can implicate CXR interpretation, Table \ref{Table: Patientfactors} summarizes key factors that  influence image quality and may introduce model biases. Amongst others we learned that portable X-rays machines, that are used on ICU, produce lower resolution images compared to radiology imaging department machines used in Stroke care. Radiographers also described practices of adding image annotations like “AP ERECT SITU” or “SUPINE ITU” and other  hospital or patient-related information directly onto the X-ray, which has additional implications for data anonymization. Further, we learned that some radiology reporting gets outsourced to external companies, which may introduce variances in reporting style or quality compared to in-house reports. In other words, any systematic differences in image resolution, image annotations/ masks, factors like patient rotation (e.g., if a proxy for other patient pathology), or any particularities in 'outsourced' reporting (e.g., high prevalence of this occurring at night) may need consideration in model development for potential risks of introducing spurious correlations. 

Lastly, poor image quality can also be a consequence of patient factors such as their \textit{level of impaired consciousness}\footnote{In types of acute medical and trauma patients level of impaired consciousness is recorded via the Patient Glasgow Coma Scale)} whereby the inability of unconscious patients to proactively hold-in their breath at image capture correlates with poorer image quality. Previously, we also described how factors like patient obesity and other chest abnormalities that show as white opacities on the image and patient conditions (e.g., history of gastric surgery, hiatus hernia or situs inversus) need consideration as they can either complicate image interpretation (by human or machine), and mean that correct placement for an NGT likely varies from more standard cases.

\begin{table*}

  \caption{Summary of patient factors and potential confounders in image or report data that can implicate AI analysis.}
  \label{Table: Patientfactors}
 
  \small
  \begin{tabular}{ll}
    \toprule
    \multicolumn{2}{l}{\textbf{Edge-cases \& Confounders for AI Data Analysis}}\\
    \midrule
    Image quality (acquisition) & Poor image penetration/ exposure (e.g., image capture settings)\\
     & Lower image resolution (e.g., type of imaging system)\\
     & Image annotations (e.g., AP ERECT SITU, SPINE ITU)\\
     & Patient rotation\\
     \midrule
    Patient factors & Impaired consciousness (e.g., Patient Glasgow Coma Scale)\\
     & Patient obesity (e.g., BMI)\\
     & Chest abnormalities that show as opacities (e.g., lower lung consolidation)\\
     & Other patient history/ conditions (e.g., gastric surgery, hiatus hernia, situs inversus)\\
    \midrule
    Radiology report quality & Differences between in-house vs. external reporting\\   
    \bottomrule
  \end{tabular}
\end{table*}  

\subsection{Opportunities and Challenges for Data Preparation \& Labelling}
This section illustrates how insights real-world practices surrounding NGT image capture, review and reporting can surface challenges for, as well as assist in processes of, effective data preparation and labelling. 

\textbf{Challenges in Effectively Linking Radiology Image(s) with Reference Text:} Much AI development in radiology imaging involves AI training on carefully curated datasets. While a radiology study may include multiple X-ray images (e.g., various frontal and/or lateral views) and their corresponding radiology report text (cf., MIMIC dataset~\cite{johnson2019mimic}), model training often requires a simpler mapping, for example in the form of 'one image - one report' or 'one image - (multiple) label' pairs. To achieve such formats by processing real-world data -- whose production may be less standardized and messy -- however can be more challenging. Next, we describe three examples of current CXR data review and documentation practices that reporting radiographers described, which we believe are relevant to data preparation efforts as they: (i) can contribute to decisions to include or exclude, i.e., incompletely imaged studies from AI analysis (example 1); and (ii) suggest greater complexity and higher chances of errors for applying, i.e., automated approaches to precisely map radiology images with report findings (example 2); or extract key entities (e.g., a sentence mentioning the NGT and its position qualifier) for data label generation purposes (example 3).  These data practices are:

\begin{enumerate}
    \item \textit{Multiple (potentially sub-optimal) radiology images can be taken to complete one CXR order.} For instance, if one CXR image misses the apices of the lungs and another image captures better the lower abdomen (crucial for NGT assessment), both images would be sent to PACS to provide “a better diagnostic picture” (RR2) to the image reader, who would mentally assembly them. 
    \item \textit{Multiple images being documented within one radiology report. }Radiographers remarked on the non-ideal, yet common reporting of image series, especially for inpatients where X-rays are taken daily. In those cases, radiographers load 3-4 images into the same accession and reports them chronologically within the same report. 
    \item \textit{Multiple lines and tubes being reported within one report sentence.} Our review of NGT data records showed how multiple patient lines and tubes that would show on the CXR would be referenced together within a single radiology report sentence (e.g., “Appropriately sited left CVC and NG on most recent radiograph”; “On the most recent image appropriately sited NG and ET tube approximately 4.5 to 5 cm above the carina”). 
\end{enumerate}

\textbf{Ambiguity in Defining \& Labelling NGT (Mis)Placement:} In addition, we found that, although NGT review was perceived as generally a straight-forward assessment task, there were differences in how participants defined if a tube is \textit{correctly} or\textit{ sub-optimally} placed. While there was agreement across clinicians that a \textit{critically misplaced }NGT was one that lead into the lung -- the condition that most existing ML models also predict for best (e.g.,~\cite{drozdov2023artificial}); assessment of correct or sub-optimal NGTs could vary in practice from text-book definitions. Clinicians described how their review was influenced by: (i) the likely implications of a sub-optimal NGT placement on the patient (a tube that may be too far down poses lower safety risks than one not advanced enough); and (ii) reporter experience and judgement on whether the NGT is likely in a good-enough position – even without clear visibility of tube tip (e.g., to not have to repeat the X-ray). Furthermore, inter-personal assessor variances in exact location/ NGT definitions as well as in their reporting styles (e.g., level of precision, interpretation, recommendations for action) can \textit{create ambiguity and variance in establishing robust NGT-positioning labels}. We illustrate this futher below.

\textbf{\textit{Correctly Placed NGTs:} }Generally, a correctly placed NGT is “in the right place in the stomach” (D3). To make assessments of correct placement, ICU doctors referred to the NGT template or similar criteria, protocols or medical guidelines/ rules that enable their assessment of whether a tube is correctly placed. However, to conform with those visual checks requires the ability to (i) \textit{fully trace the path of the NGT}; and, ideally, to (ii) see the \textit{tube tip}. While one ICU consultant expressed lesser need to see the actual tip (“I know that's in the GI tract probably rather than necessarily see the tip”, D6), most described how difficulties to locate the tip would cause hesitation in decision making given the high-stakes implications of any mistakes. A Stroke registrar explains the judgement call that is then made:

\begin{quote}
\small
    “It's quite common for either the problem to be it's in the GI tract, but the tip is so far that it's actually off the bottom of the X-ray. (…). And in that situation there's a bit of a judgment call to be made of: Do you think the tube is in the right place, but the X-ray didn't go far enough? Or do you think the tube is too far in? And in those situations either you might say, ask the nurse to pull the tube back a few centimetres, up to five centimetres. Try and re-aspirate. And if they can't re-aspirate repeat the Chest X-ray. Or I just call the radiographer saying the tube looks like it is definitely in the right place. I think that X-ray just didn't go quite far enough, could you repeat that X-ray just going down a little tiny bit further? That's quite common.” (D7, Stroke)
\end{quote}

\textbf{\textit{Sub-optimally Placed NGTs:} }For sub-optimally placed NGTs, participants distinguished between three types: (1) the NGT is 'folded up', 'bend' or 'coiled' within the patient’s mouth, oesophagus or stomach, which can obstruct feed to pass through and increase risks of lung aspiration; (2) the NGT does \textit{not reach far enough into the stomach}; or, more rarely, (3) is \textit{inserted too far in}. In those two latter cases, all reporting radiographers and some clinicians described difficulties and judgement calls to determine exactly how far the NGT would need to be advanced for it to overall be considered still safe for feeding, and when to alert clinicians that its exact position is not within the stomach: 

\begin{quote}
\small
    “(…) one of the things that is hard to bottom out is: at what point in the stomach is it safe to feed? Because if it's at the gastroesophageal junction, do you run the risk that any patient head movement will then dislodge the tube that becomes oesophageal? And how do you say, you know, it’s 5 centimeters past the GOJ\footnote{The GOJ (gastro-oesophageal junction) is the part of the gastrointestinal tract where esophagus and stomach are joined. It's a key landmark to assess if the NGT extends far enough and reaches into the stomach. However, the GOJ is not directly visible on a Chest X-ray}? Part of the problem about standardizing the interpretation; expectation and assessment for radiographers is, is it 5 or 10 centimeters passed the GOJ, because once you've set that standard, that's it. For me, as long as it appears radiographically clear of the region of the GOJ, obviously the Chest X-ray is 2D flat. You know about where the GOJ is, if it is over the stomach. It's safe to feed. It's OK.” (RR1, ID)
\end{quote}

Some ambiguity how individual reporters exactly define or distinguish what counts as safe NGT placement is also reflected in reporting practices. One radiographer remarks that some reporters “consider sort of 5cm past the gastroesophageal junction as adequate, other people say you need to advance it” (RR3); commenting on report variations:

\begin{quote}
\small
    “I guess the other challenge is you know the variability to people's opinions and I guess variability of what people would say is an optimal position. What are the guidance they're gonna give, what advice they're gonna give in their report as well, cause a lot of people would just say satisfactory, something like that, or advanced, but they won't go into specifics, so there's that variation of reporting.” (RR3, ID)
\end{quote}

\hfill \break
\textbf{Leveraging Standardized, Vision-based Reporting Practices: }Identified variations in clinical assessments likely complicate definitions of a robust ground truth for model training or performance evaluations. One approach forward may be to again take inspiration from existing clinical practice. To manage uncertainty in image interpretation, ICU clinicians and reporting radiographers described referring back to the .NGT template or similar protocols, and to revert to describing the tube’s location rather than making any safe to feed decisions or other recommendations. Aligning with the provision of clear visual descriptions that the clinical team can then interpret in the context of an individual patients’ circumstance may also present a less risky, potentially clinically more useful AI output than, for instance, more binary, high-level NGT classifications (e.g., confirm correct placement). In this regard, the .NGT template (Figure \ref{fig:template}) offers a unique opportunity for framing AI tasks more specifically around visual assessments. Not only does it provide a focus on visual checks of the feeding tube with regards to the CXR (e.g., “The NGT bisects the carina”; “The NGT pass below the diaphragm”); its standardized format that records yes/ no responses to specific visual questions also makes it particularly suitable for auto-label extraction; and offers concrete criteria to evaluate AI models against. 

\subsection{Opportunities \& Challenges for Measuring AI-Intervention Success}
In this final section, we engage with the question of what would be relevant outcome metrics for assessing the effectiveness of an AI NGT application if deployed into practice, as well as the practical feasibility and reliability of assessing such metrics giving existing data documentations. Table \ref{Table: Outcomes} summarizes potential outcomes of interest with regards to AI-assisted feeding tube (mis)placement verification. 

\begin{table*}

  \caption{Summary of potential outcome metrics for assessing improvements to feeding tube placement assessment, and timelier verification or use with goals to improve patient safety or workflow efficiency.}
  \label{Table: Outcomes}
 
  \small
  \begin{tabular}{ll}
    \toprule
    \multicolumn{2}{l}{\textbf{NGT AI: Outcomes of Interest}}\\
    \midrule
    Improved patient safety & Reduced number of Never Events (e.g., as indicated in DATIX records)\\
     & Reduced number of misplacement-induced complications\\
     & Reduced number of missed NGT misplacements\\
     & Reduced time to identification of misplaced NGT \\
     & Reduced time to documentation of misplaced NGT (e.g., in radiology report/ clinical note)\\
     & Reduced time to correction of misplaced NGT (e.g., NGT removal/ replacement)\\
     & Reduced errors in image assessment\\
     \midrule
    Improved workflow efficiency & Increased reporter confidence\\
     & Reduced requests for human peer review\\
     & Reduced time to patient feeding\\
    \bottomrule
  \end{tabular}
\end{table*}

\textbf{Improved Patient Safety.} A key motivator to the application of AI is to help improve patient safety by reducing risks of sub-optimal or critically placed NGTs being used for feeding. This may be achieved: (i) by reducing occurrences of NGT misplacement remaining undetected; or (ii) by speeding up their detection.

\textit{Reducing undetected misplacements: }Perhaps the most obvious metric for patient safety is to evidence a reduction in number of Never Events, which are officially reported via DATIX – a risk management information system designed to collect and manage data on adverse events~\cite{TheMedicalEducationDirectorate}. Whilst highly significant for patient safety, the extremely low prevalence of such events can mean it is too sparse to serve as reliable metric - at least for any smaller-scale, short-term pilot study. NGT misplacement into the lungs without feeding, will likely occur more frequently, and requires as early detection as possible. Here, EHR data may give useful context to a patients’ physiological responses either via documented CEASE\footnote{CEASE signs stand for: Coughing, Extreme agitation, Abdominal distension, Stoma site leakage and Elevated temperature. These can all indicate complications that arise from feeding tubes and can be recorded in the patient record. } signs or detected increases in heart rate or drops in oxygen saturation levels that can be picked-up at the time of NGT placement, and can be documented within EHR ICU data. It may also be worth assessing in any subsequent patient X-rays (e.g., using AI), if signs of lung aspiration surfaced that could be indicative that an NGT misplacement had been missed in previous verification steps (e.g., deceptive aspirate test, of CXR image interpretation error). Technically, a reduction in image assessment errors also presents a useful metric to evidence AI success, however, as stated previously, it is currently unclear how prevalent image interpretation errors are in NGT CXR checks as these may not be spotted or formally documented (outside Never Events).

\textit{Speeding up misplacement detection:} An early alert to a misplaced NGT to clinicians, or to re-prioritize reading lists is sought to speed up the time of its detection. To trace specifically the \textit{time that a misplaced NGTs has been detected}, in data, however, can be more complicated (e.g., a radiographer may spot a misplacement already at time of image acquisition). A more feasible and reliable metric may be earlier documentation of misplacement through a faster radiology reporting turn-around for misplacement tubes or quicker clinical documentation. Ideally, as a consequence of an early misplacement detection, time to its correction should be reduced. This may further be evidenced in data entries documenting the tube removal and reinsertion of an NGT post CXR (e.g., on a patient avatar within EPIC documentations); and time-stamped clinical notes describing corrective actions. 

\hfill \break
\textbf{Improved Workflow Efficiency: }Improvements to workflow efficiency can be harder to capture or evidence. Outcome metrics of interest may be (i) increases to staff confidence in image interpretation and therefore reduced need for peer review due to AI assistance; and (ii) speedier NGT correction and patient feeding.

\textit{More confident, speedier image review: }AI proposals that may notify about potential human reporting errors or that improve visual image review (e.g., via an overlay), suggest to increase staff confidence in image interpretation, which could be qualitatively assessed and may be indicated through proxies such as a reduced need for peer review as well as speedier review times.

\textit{Reduced delays to patient feeding: }As a consequence of more effective misplacement detection and correction, it is assumed that AI assistance can help reduce delays to patient feeding. While this may be the case, we need to consider for this and other timing related outcome metrics: (i) how accurately they are recorded; and (ii) what the opportunities are for staff to act upon earlier alerts and review requests. For example, machine generated data, image order requests, their upload and signed documentations as well as feeding records provide more accurate timing information than, i.e., a nursing notes that get composed and added to throughout the duration of their shift before it is signed and time-stamped at the end of that shift. Furthermore, it could be argued that (ICU) doctors already look at the NGT CXR at their earliest convenience; and – as described earlier – we need to account for the time dependencies based on other ward dynamics (e.g., shift rhythms, emergencies, staff resourcing, time of day), or changes in plans to feed the patient (e.g., patient may undergo surgery first). This suggests speeding-up the detection of an NGT misplacement as best captured by the timing of its documentation, which likely presents a more reliable outcome metric than subsequent clinical action of patient feeding. 

In summary, this section highlighted how additional data may need to be included in evaluations for key outcomes of interest (e.g., DATIX data, patient complications at time of NGT placement, or follow-up CXRs to assess for potentially missed sub-optimal NGTs); and suggests a more robust and reliable assessment of AI effectiveness via data items that require mandatory documentation; are set-up to be documented at specific times or time-intervals (e.g., 4-hourly aspirates or regular feeding logs); and that are less time-dependent on broader hospital dynamics. 

\section{Discussion: Context-Specificity \& Complexities for Realizing AI Utility}
Much of current AI development has been criticized for being technology-driven, decontextualized from concrete use scenarios and the many human and organisational factors that affect their adoption and integration within healthcare practice (e.g.,~\cite{liao2022connecting, miller2017explainable, thieme2023foundation}. To better address the disconnect between technical AI capabilities and real-world stakeholder needs ~\cite{coiera2019last, osman2021realizing, zajkac2023clinician}, this paper contributes an in-depth case study describing our approach and the learnings of bringing a human-centered process to early stage AI innovation work that seeks to understand current clinical practice and identify the right problems for AI to solve. Next, we discuss the implication of this work that (i) surfaced complex interrelations between human, technical and organisational factors that determine perceived AI utility; (ii) discuss challenges surrounding human expectations of AI and configurations of human-AI interactions for fostering AI acceptance; and (iii) reflect how key insights into real-world data production and its characteristics can usefully guide AI development processes. Across these areas, we draw out directions for future work in healthcare AI, and for radiology more specifically. We conclude with some of the (iv) limitations of our work. 

\subsection{Interrelations in Perceived AI Utility}
By grounding our AI work within the specific use context of an ICU hospital and concrete NGT CXR verification workflow, we learned how perceived clinical utility of AI capabilities and their potential realization within healthcare practice are bound-up by a complex interplay of multiple factors. In this section, we draw out how identified factors of AI goals and implications; workflow design and use integration; real-world data production and actual technical realization of AI capabilities are interlinked; and how additional considerations of broader medical-legal, IT infrastructure, and resource constraints determine trade-offs in benefit/ cost relations that implicate perceived AI utility. We suggest activities of systematic mapping as tools to clarify those relations and facilitate (cross-disciplinary) discussions and decisions on directions forward.

\subsubsection{Trading-Off Intended Use with Actionability, Data Availability and Broader Organizational Constraints} 

\hfill \break
When reviewing our five proposals for how AI could assist NGT CXR practices, we identified desires for patient safety as a predominant driver of clinical relevance. While goals for patient safety can be assisted by efforts to increase staff confidence in image assessment, and by speeding up overall processes (e.g., reducing delays to misplaced NGT detection and correction); AI proposals that more concretely pronounced \textit{patient benefit and safety} received most support. This suggests a focus on AI that facilitates \textit{more immediate detection of critical findings}, or \textit{human errors}. When we investigated where and how within existing workflows such AI functionality could be clinically most relevant and impactful (e.g., to speed-up and inform actual care decisions) our research surfaced, amongst others,  interdependencies with factors such as: (i) \textit{timing} and \textit{staff’s ability to act} upon AI insights; as well as (ii) broader clinical care pathway and organisational set-up considerations; and (iii) data availability constraints.

\textbf{\textit{Ability to Act within the Context of Existing Workflows and other Hospital Dynamics. }}
Investigating different workflow stages (illustrated in Figure \ref{fig:workflow}), we learned about the roles of various stakeholder groups involved (as potential AI users) and how the timing of an AI interventions interlinks with clinical utility. For example, having an AI potentially alert to any critical or sub-optimal NGT placement detection ideally \textit{as early as the image acquisition stage} would be most beneficial. This workflow stage involves imaging radiographers, who were also identified as the only constant across care settings, suggesting greater potential for an AI application to \textit{scale} and expand reach of benefits. Yet, imaging radiographers current \textit{professional qualifications} and \textit{medical responsibilities}, mean that they can only inform referring clinicians/ nurses about any noteworthy NGT observation, but often do not have permission to report the NGT CXR. This workflow integration proposal may further be complicated by technical constraints such as requirements to have AI run on X-ray machinery. In contrast, an AI that would, for example, support 'human error detection' in the reports of radiology reporters (radiologists, radiographers) may be technically easier to integrate within reporting software (e.g., PACS), but likely has a lesser impact on clinical practice, since ICU doctors – by the time the CXR gets officially reported, usually multiple days after CXR capture – will have already assessed and acted upon the NGT image, This renders the official report ‘irrelevant’ – whether it was correct, or not. As a consequence, such AI functionality will only unlock its utility if combined, i.e., with image review prioritization functionality. 

The latter example hints at the importance of staff being able to act upon AI insights, ideally in ways that improves current processes/ patient outcomes, for AI utility to become realized; as is also discussed in other research~\cite{lindsell2020action, zajkac2023clinician}. Staff role responsibilities and competencies aside, our study further showed how \textit{resource availability, shift patterns and broader ward dynamics (like emergencies)} influence staff’s ability and decisions to complete, or hold-off on NGT specific tasks. Given these constraints and that ICU clinicians described to already try and review NGT CXRs as early as possible; or that reporting radiographers, for example, may not be available out-of-hours; this surfaces the question, to what extent, realistically, staff could 'better action' any early alerts or worklist prioritization in response to a critical findings detection. At a minimum, our findings suggest that key context variables such as staffing levels; shift and work hours; or prevalence of emergencies may need careful consideration in any evaluative study protocols that seek to assess AI effectiveness in practice. In additional to work rhythms, research by Beede et al.~\cite{beede2020human} draws attention to considerations of the \textit{built environment}. Their work showed, how poor clinic light conditions at a local clinic negatively implicated the performance of their AI system that otherwise showed high accuracy in lab tests.

\textbf{\textit{Broader Clinical Care-Pathway \& Organizational Set-up Considerations. }}Furthermore, we need to consider the development of any one AI application in the context of the broader organisational set-up. For one, our work surfaced how textit{care pathways varied} between ICU and Stroke services, which was evident amongst others in differences in: (i) image capture technology that can mean different software integration requirements, or image quality outputs; and (ii) role responsibilities that suggest a greater reliance on official radiology reporting in Stroke care. These suggest potentially the need for tailoring solutions and configurations of AI to different pathways. Further, we need to account that sub-optimally placed NGTs, of course, present only one of many potentially ‘urgent’ findings that could be detected and prioritized on a CXR. For example, Seah et al. ~\cite{seah2021effect} list 34 ‘critical’ findings to detect on CXRs, and that could compete for clinical attention. Furthermore, research by See et al.~\cite{see2023alerts} shows that – even outside of any AI use – and despite the availability of a well-established electronic notification systems and mechanisms to notify referring clinicians about an abnormal radiology report, urgent findings can still be missed or their communication be delayed; arguing for the important to have \textit{the right organisational set-up to be able to translate important clinical findings into prompt actions} (cf.~\cite{Elish2020Repairing} -- above and beyond mechanisms to detect or alert to critical findings.

\textbf{\textit{Balancing Data Availability Constraints with Broader Opportunities for AI Innovation \& Impact.}} Data availability limitations can affect the feasibility and evaluation of some AI solutions. For instance, for AI to detect reporting errors in image assessment, it is unclear how frequent these errors are, as they may go unnoticed or unreported in historical data, or are only documented in critical incident reports when not all initial errors lead to critical incidents. This makes it hard to obtain and process such data for AI purposes. While data, its quality and scale, necessarily presents the key building block for the practical realization of any AI, it is important that this does not necessarily limit explorations of other potentially more viable use cases and areas of AI opportunity (cf.~\cite{thieme2020machine}). While our research investigation and AI proposals specifically focused on addressing NGT CXR image assessment or report generation challenges, our study surfaced ‘delays’ that spanned across the entire NGT verification process as the most prevalent workflow problem. This suggest \textit{broader opportunities} to assist in clinical process optimizations. For example, our work highlighted communication inefficiencies whereby staff described the constant checking and chasing-up of tasks; as well as tedious efforts to extract relevant patient information from EHR records (cf.~\cite{sahniadministrative} for a report on costs and need to simplify administrative burden in healthcare). This also foregrounds how definitions of desired AI functionality may best evolve as an iterative dialogue between user need and data availability. 

\subsubsection{Systematic Mapping as a Tool for Achieving Clarity about Key Factors and their Interrelations}
Given the complexity of the various human, data and organisational factors that implicated perceived AI utility, we found it helpful to systematically map out key factors and their interrelations across different AI proposals. We illustrated an excerpt of such mapping in Table \ref{Table: Map}, which depicts links between: workflow stages, users, design choices, AI error type breakdowns, direct and indirect stakeholder implications, and broader technical or medical-legal constraints. We believe that this –  or similar activities (cf., Yang et al.’s~\cite{yang2020re} AI design complexity map; or design resources for scaffolding AI concept ideation by Yildirim et al.~\cite{yildirim2023creating}) – can serve as a useful exercise to trade-off perceived AI benefits and risks, costs and feasibility constraints and thereby provide guidance to research and development teams to make more informed choices on AI use cases and their configuration. Insights revealed will likely provide valuable inputs to current responsible AI practices such as impact assessments~\cite{ assessment2022case,MicrosoftRAI}, and serve as boundary objects~\cite{john2004identifying, lee2007boundary} - as common frames of reference to facilitate inter-disciplinary team collaboration (cf.~\cite{ayobi2023computational, cai2021onboarding}) – when discussing i.e., clinical priorities; realistic AI performance goals, or necessary risk mitigations.

\subsection{Human-Process Integration of AI}
This section discusses how perception of AI utility and acceptance of future AI applications are bound-up with often high-expectations of AI performance, and how AI functionality becomes positioned and integrated within human work.

\subsubsection{Setting \& Managing Appropriate AI Expectations}
Our study findings surfaced how participants evaluations of perceived AI utility was based on how well it met their expectations of its performance. For AI proposals like NGT report generation, to realize its full potential, they suggested the AI needed to be high-performing, ideally 100\% accurate. More value was also ascribed to AI that would surpass human capabilities – particularly for assistance with complex and difficult-to-interpret patient cases. However, it is unclear how well AI would handle edge-cases; and how realistic an almost perfect AI performance is. Other research~\cite{matthiesen2021clinician, zajkac2023clinician} has also warned that unrealistic and overly high expectations of AI can negatively affect its perceived usefulness. We also noticed a tension between aspirations and excitement for how AI could be transformative to staff practices or patient care (e.g., references to ChatGPT capabilities), and staffs previous experiences with other AI, decision-support or automated systems (e.g., auto-generated ECG impression) that made them doubt AI capabilities. This tension made more ambitious AI proposals seem more of a theoretical proposition rather than practical reality. Zaj{\k{a}}c et al.~\cite{zajkac2023clinician} also found that there is more familiarity with AI use for decision support (including quality assurance) and prioritization, than with AI for automation scenarios, which are still very rare in healthcare  (cf.,~\cite{beede2020human} for an exception).  It is a well-recognized challenge in human-centred AI design to help healthcare professionals and other non-AI domain experts to \textit{develop realistic expectations of (new) AI capabilities} (e.g.,~\cite{yang2020re, yildirim2023creating}). This implies a need for better understanding of how healthcare professionals currently perceive AI (cf. ~\cite{petitgand2020investigating}), as well as for more support of cross-disciplinary and education initiatives that enhance healthcare stakeholders' knowledge of AI. This will enable a more informed perspective and better participation in discussions on feasible goals and strategies for AI (system) innovation -- both short- and long-term. 

\subsubsection{Appropriately Configuring Human-AI Relations}
Whilst participants described the potentially benefits (e.g., time-savings) of an AI to automatically assesses the CXR image and generates the NGT report; to unlock most utility, they also described for the AI needing to take ‘full responsibility’ of its outputs, alongside requirements for high performance and current medical-legal barriers. Where, instead of full automation, AI functionality was positioned to only assist existing human practices (e.g., by showing an image overlay to help visual analysis, or creating a ‘preliminary’ report) it was harder for staff to identify the value added by AI (cf. findings by~\cite{yildirim2024multimodal}). Clinicians would still have to do their own checks according to current practice, and would be fully accountable for their medical judgment. Suggested benefits of the AI would also be weighed-off with AI risks. This includes potential negative effects on staff workflows and patient outcomes due to AI errors (e.g., increasing peer review when false outputs cause confusion), and risks of AI over-reliance; or conversely, of AI alert fatigue, if its outputs were incorrect often. Proposed AI benefit also needs to be high enough for staff (and hospital organizations) to be willing to invest time and effort (e.g., staff training) to potentially adjust work practices to accommodate new AI functionality; along with required IT infrastructure and other financial resources and information governance processes involved in developing healthcare AI services~\cite{rees2023information}. This includes considerations of opportunity cost whereby resources are expended in one area, but not another~\cite{lindsell2020action}. Thus, given the considerable resources that are needed to successfully build and deploy AI systems in healthcare, it is important to understand the conditions under which AI can be effectively leveraged to maximize the benefits these investments can bring to healthcare delivery~\cite{petitgand2020investigating, sendak2020real}. Even where research and development demonstrate clear clinical benefits, balanced with costs of disruption or change, there still needs to be willingness to pay for new (AI-enabled) services. 

\subsubsection{Shifting from ‘Human-AI-Verification’ to ‘Human Process Integration of AI’}
Where effective, responsible realization of (full) AI automation – depending on the use case – may present a longer-term ambition, we speculate that greater utility may come from: a better integration of AI within healthcare work; and shifts in clinical role responsibilities.

\textbf{Limitations of Current Strategies to De-Bias and De-Risk Potential AI-Errors.} Our research findings echo previous studies that surface tensions in\textit{ human-assisted} or \textit{human-in-the-loop AI} systems to balance desires for a seamless, unobtrusive AI integration (e.g., \cite{thieme2023designing, yang2019unremarkable}) with approaches to ensure humans appropriately interact with AI outputs by taking time to check its correctness. Especially in healthcare, where clinicians work under time-constraints, they may not have the interest, ability, nor technical expertise to engage more deeply, or more critically with AI outputs~\cite{jacobs2021designing, sendak2020human}. When extra time is spent on data review, waiting for AI outputs, or sorting through (false) AI alerts, this can disrupt work practices and distract clinicians from their focus on patient care~\cite{sendak2020real, zajkac2023clinician}. Similar concerns are reported in recent research by Bach et al.~\cite{bach2023if}, who specifically study bias mitigations in clinical AI support by asking humans: to make their assessment first; to provide decision justification; or to explicitly consider opposing AI outputs. Their study showed that while such strategies can reduce bias and improve diagnostic accuracy, the \textit{additional burden required to engage with AI in this way}, significantly decreased work efficiency. Clinicians also did not appreciate for AI to correct rather than support them in their tasks – describing the experience as condescending. All this suggests a more complex picture when trying to appropriately configure human-AI experiences that are \textit{seamless, effective} and that \textit{responsibly} consider the nuances of different AI error scenarios.

\textbf{Taking Inspiration from Existing Safeguarding Practices. }For a safe, effective integration of AI outputs within clinical work, we therefore wonder if more inspiration could be taken from existing hospital safeguarding practices designed to exercise appropriate caution and prevent human errors. Amongst the range of risk mitigation strategies we discovered were: (i) a clear hospital NGT policy, staff training and skills test; (ii) the cultivation of a cautionary mindset; (iii) human peer review; (iii) mandatory, templated reporting of the NGT; and (iv) continuous (4-hourly) position and aspiration checks. These strategies and their combination, whilst not in themselves perfect, are designed to catch human errors. Rather than positioning humans as “error-checker for (imperfect) AI”, maybe the aim should be to extend investigations how AI insights can become part of familiar steps and processes of clinical information review, peer checks, guideline adherence, and other practices to clinically correlate and validate medical insights. How might a framing of AI insights within such practices and information ecosystems aid its utility and acceptance? Treating AI insights pragmatically alike other data tools that offer unique insights and have their limitations, the goal for users should be less on identifying whether a specific AI output is correct or not (or how it arrived at a particular output/ prediction). Instead, we should aid  clinicians to effectively triangulate AI insights with other evidence and information they have about the patient such that it brings about caution or confidence in taking next steps. This suggests connecting AI outputs to other criteria ‘external’ to the model~\cite{shanahan2022talking, thieme2023designing} and its bringing into context with other, trustworthy information agents or resources (cf.~\cite{yang2023harnessing}) to \textit{facilitate meaningful clinical correlations that enable clinicians to either accept or reject/ ignore AI outputs similarly to how they would treat other ‘imperfect’ assessments}. This suggests for future work to explore more deeply the integration of AI insights within existing (safeguarding) processes; alongside continued advances in technical solutions (e.g., self-consistency prompting~\cite{singhal2023large}, LLM-generated explanations~\cite{nori2023capabilities}, or correctness predictions ~\cite{kadavath2022language}) that serve to reduce (risks of) AI errors. 

\textbf{Adapting AI to Different Users \& Supporting Shifting Role Responsibilities. }To expand reach of radiology services and reduce the implications of delayed official radiology reports, we learned how all ICU clinicians are trained and permitted to assess NGT CXRs; and about the role of reporting radiographers – including near-term plans to upskill imaging radiographers to be able to make all-important safe-to-feed decisions. Such shifts in roles and responsibilities expands the workforce that takes on more radiology-specific tasks, which may indeed be necessary to increase reporting capacity required to address ever growing imaging backlogs~\cite{maskell2022does}. On the one hand, this suggests \textit{AI may need to be adapted for different user types and their needs}. On the other hand, this surfaces the question: \textit{how AI could play a useful role in building-up image assessment confidence without getting in the way of important human skills-acquisition}? Here, tasks like CXR NGT verification may be particularly suited to AI support as it has a clear question (e.g., is the tube in the correct place?) and involves an observation-based assessment rather than a complex clinical interpretation. Another example may be: is there a bone fracture? For these types of tasks, visual inspection following standard, protocolled procedures is sufficient and \textit{does not demand more advanced, expert-level radiology evaluation}, as is often the case for higher-risk AI tasks that involve: diagnosis, treatment suggestions, and other decision-support functions (cf. ~\cite{yildirim2024multimodal}). Whilst this suggests opportunities to leverage a wider workforce - at least for certain radiology tasks -- this is not without challenges, as a CXR may reveal other medical conditions that require clinical intervention; raising questions about diagnostic boundaries and medical-legal responsibilities that need to be addressed in future work.

\subsection{Relevance of Real-World Insights into Data Production and its Characteristics for AI Development}
The last section of our findings details learnings about existing NGT hospital data -- it's production, availability, and characteristics – and discuss their relevance in guiding AI development specific to NGTs on CXRs, and beyond. 

\subsubsection{Implications for Data Preparation \& Model Training/ Evaluation}
\hfill \break
\textbf{Accounting for Potential Data Biases \& Including Relevant Context Data for Image Interpretation.} Through our contextual inquiry and interview research, we identified key factors that can influence image quality, introduce model biases, or otherwise complicate image interpretation. For example, we described how key patient factors such as obesity, conscious impairedness, and obscuring structures or pathologies their chest can hinder the visibility of the NGT path or tip, whilst edge cases such as an unusual or changed patient anatomy can mean deviations in image interpretation (e.g., what’s considered ‘correct’ placement) from more standard cases. This suggests the inclusion of key image meta and EHR patient data (e.g., patient BMI, Patient Glasgow Scale, patient medical history) as important context information to image assessment and AI model training/ testing. 

Our findings also identified specific data characteristics that can invite spurious correlations that may need controlling for. For instance, specific image makers that we observed in CXR NGT data, such as “AP ERECT SITU” or “SUPINE ITU” annotations, whilst not entailing any patient-identifiable information that otherwise would be removed, can be indicative of 'ICU' patient care. Even where those annotations are masked (e.g., using a black box), their visual remains and specific locations can become spurious ‘shortcuts’ a model might learn instead of desired image features (cf.~\cite{degrave2021ai}). Here, new methods such as RadEdit~\cite{perez2023radedit}, a generative image editing approach may be particularly promising, as it allows to systematically mask image parts to diagnose potential spurious correlations and other biases. 

As another example on the risks of spurious correlations, Drozdov et al.~\cite{drozdov2023artificial}, who developed a deep learning model to classify NGT position, found their model performance to be adversely impacted and the model to be very sensitive to images that showed sudden changes in system manufacturer and institutional department, which assumes differences in imaging machinery used, and resulting image quality. This suggests the inclusion of different hospital departments or imaging system details within the data strategy; alongside attributes of patient age, sex, or ethnicity that are more commonly considered to ensure a diverse, representative dataset composition~\cite{ueda2023fairness}, and to assess AI fairness~\cite{ahmad2020fairness, mbakwe2023fairness}. 

\textbf{Understanding Data Production for Effective Data Curation}. 
In Section 5.5, we described how real-world NGT reporting practices can implicate effective dataset curation. For example, in much AI analysis that uses prominent Chest X-ray datasets (e.g., MIMIC~\cite{johnson2019mimic}), NIH ChestX-ray14~\cite{wang2017chestx}, PadChest~\cite{bustos2020padchest}, Indiana CXR collection~\cite{demner2016preparing}); and also more lines and tubes specific ones (e.g., RANZCR CLiP~\cite{tang2021clip}), data is often represented as “image – report" or "image – (multi) label" pairs. Achieving a simple mapping between data inputs (e.g., 1 image - 1 report sentence) may be harder, where such information needs to be disambiguated from text reports that address multiple images, or where multiple images are linked to one report. These are important considerations where label generation as part of imaging dataset creation increasingly uses more automated methods (cf. CheXpert~\cite{irvin2019chexpert}, PadChest~\cite{bustos2020padchest}) that utilize advanced natural-language processing capabilities to  identify and extract relevant text entities from radiology reports in-lieu of requiring expensive 'human' label annotations. Fortunately, recent advances in large-language models (LLMs), and their combination with careful 'prompt' strategies, demonstrate that entity extraction from radiology report texts can be very effective for a range of tasks, even if only few examples are given to the LLM (cf. ~\cite{liu2023exploring}). 

\textbf{Complex Data Inclusion/ Exclusion Decisions.} Our study findings also surfaced common challenges of CXRs potentially omitting key parts of the relevant patient anatomy, e.g., cropped the lower abdomen or the apices of the lungs are missing. With current trends in healthcare AI that focus on improving data quality in training (e.g.,~\cite{he2019data, kjelle2022assessment, liu2023exploring}), this might suggest excluding such studies from analysis. To a certain extent this \textit{data exclusion may be necessary to ensure robust AI training} -- given that the relevant area for the target AI task might not (fully) show on the image. Moreover, when deploying the AI system, \textit{it is important to define what kind of inputs are acceptable for the model to produce reliable and safe outcome}s. Simultaneously, however, this data exclusion can mean that a substantial amount of imaging studies could be rejected for analysis, limiting the potential reach and perceived utility of a resulting AI application. For example, in the context of the RANZCR CLiP dataset~\cite{tang2021clip}, 33\% of NGT images (2748 out of 8344) were labelled as "incompletely imaged", which constitutes a considerable amount of studies that could risk not being predicted for alongside other \textit{data exclusions} (e.g., restrictions to adult populations).  

\textbf{Defining Data Labels/Outcomes \& Leveraging Existing, More Standardize Clinical Data}. Interlinked, our study surfaced how there can be ambiguity and variance in cases where assessment of a correct or sub-optimal NGT presents a more borderline case. Here, we identified reporters making judgement calls that are guided by (i) their experience and considerations of the (ii) likely implications of errors in those judgements (e.g., it is less problematic if a tube is slightly too far advanced vs. not far enough extended into the stomach). The clinical reasoning they apply may differ from common text-book definitions that often suggest for the NG tube tip to be “at least 10 cm below the gastroesophageal junction”, a definition often used in NGT data label generation efforts~\cite{drozdov2023artificial, tang2021clip}. We also learned about differences in levels of precision or style of reporting; and potentially between in-house and outsourced reports. There are different strategies in how label ambiguity can be addressed in practice. While some datasets include labels generated or confirmed by a human expert (e.g., ~\cite{bannur2023learning, humedical}); others employ multiple experts to generate annotations and assess agreement across them to achieve more certainty about the accuracy and overall quality of the resulting labels (e.g., referred to as 'gold’ labels ~\cite{boecking2022making, }). However, such labelling efforts often require substantial resources as well as access to domain experts, which limits possibilities of their development. One route forward may also be presented by \textit{leveraging more standardize clinical notes where available}. In discovering the ‘.NGT’ reporting template, we identified not only a useful short-cut for label extraction; as a clear, standardized checklist of what the AI needs to achieve to provide utility specific to the NGT verification task, the template itself may be regarded as defining the capabilities that the AI models needs to be trained for, and to provide clear indications of assessment/ success criteria to evaluate a model against. 

\subsubsection{Evaluating AI Effectiveness (in Practice)}
Lastly, we clarified different outcomes that could be utilized to assess the effectiveness of an AI NGT application; guided by clinical goals to improve patient safety and workflow efficiency. Our data investigations surfaced how choices of suitable metrics depending on: \textit{ease of data availability and access} (e.g., routinely collected EHR data vs. DATIX data vs. data needs not-directly captured); the \textit{reliability of (timely) data capture}; especially where the goal is to assess temporal relations (e.g., to speed-up detection of critical findings); and the \textit{possibility for AI outcomes to affect change} in the context of other hospital dynamics. Furthermore, our data investigations surfaced new opportunities for analysis such as: detecting patient’s physiological responses during NGT placement (e.g., CEASE signs, detected increases in heart rate, or reductions in oxygen levels) that are recorded in EHR data as potential inputs to NGT misplacement detection; and an analysis of follow-up CXRs for feed-induced ‘lung aspiration’ to detect a potentially missed NGT misplacement or un-intentional changes to its position; expanding the scope for future work. 

Aside from considerations of additional data features, clinical outcomes, and  (automatically-computed) AI accuracy metrics, our mapping example in Section 5.3 also brought attention to the need for future work to develop more established methods to qualitatively explore AI failure cases (e.g., with domain experts); and help formalize and systematically assess the (likely) clinical implications of different AI error types. This will nuance aspects of AI models that perform better or need improving; clarify more or less permissible AI error types (based on likely implications); and help concertize definitions of desired/ successful AI performance/ outcomes within development teams.

\subsection{Study Limitations}
Our study is limited by its specific research context, chosen study method, and selected sample. 

We conducted our research within a UK hospital that is well known for its medical research and excellence in acute and specialist services. It is one of the few places that employs EPIC as an EHR system, which usually entails a substantial financial commitment for larger hospitals. The hospital also has advanced digital radiology workflows and funding for this research, indicating the availability of resources and technological infrastructure to facilitate AI innovation. We recognize that this represents a more advantaged healthcare situation. Moreover, we also acknowledge that our specific focus on ICU workflows limits the applicability of our findings to other departments, and other hospitals within and outside the UK context.

Our work is also limited by its focus on early-stage AI use case exploration and development requirements elicitation that prioritize an understanding of existing work practices and feedback on prospective ideas for how AI could assist NGT processes. As a result, we did not yet create or utilize concrete visual user interface sketches (cf. ~\cite{burgess2023healthcare, thieme2023designing, yang2019unremarkable}) nor any functional AI prototypes (cf. ~\cite{valencia2023less, xie2020chexplain}). Instead, our research insights seek to inform future human-centered AI work (e.g., key questions to ask early within research and development) and guides design concepts within the context of NGT CXRs, and healthcare AI more broadly (e.g., learnings about AI acceptance barriers, interaction design choices). As such, this work presents the starting point of a learning journey how existing healthcare data, models, users, workflows and organizations inter-relate and would have to come together to meaningfully identify, shape and realize AI innovation use cases. Future work will need to build on these initial insights through situated, iterative AI (prototype) design and development cycles.

Finally, our study sample reflects a breadth of participants rather than a more representative sample of one specific clinical user group. Jointly, with a research team also comprising experts in compliance, research governance and IT – alongside AI, HCI and healthcare, this inclusion of a broader range of professional expertise is more common for innovation studies that seek lab-to-clinic transition~\cite{zajkac2023clinician}. Even though our research team included two consultant radiologists (MTW + JJ), who assisted in designing the study protocol and interpreting the research findings, our main focus and participant selection on the ICU excluded radiologists as a user group in our study. Lastly, we also recognize the potential value of incorporating patient perspectives, for instance through more extensive public and patient involvement (PPI) beyond ethical approval, into this and future work.

\section{Conclusion}
Seeking to pave the way forward in closing the gap between innovative AI research and its translation into healthcare practice, we presented a detailed case study of how we adapted a human-centered approach to early stage AI innovation in radiology. An in-depth contextual inquiry and interviews with 15 clinical stakeholders revealed rich insights into current clinical practices, highlighting existing workflow challenges, safeguarding mechanisms and their limitations. Evaluating different AI proposals in this context, we discussed how multiple factors and complexities, such as the timing and ability to act upon AI insights, the broader clinical care pathway and organizational set-up, real-world data production and technical realization of AI capabilities as well as human expectations and interactions with AI, affect the perceived AI utility and acceptance. We proposed using systematic mapping as a tool to clarify the trade-offs and interrelations among these factors. Moreover, we argued that in configuring human-AI relations to shift from a focus on humans needing to verify AI towards a closer position of AI as integrated within existing human (safeguarding) processes of clinical information review, guideline adherence and concerns for patient safety. This brings AI as a data insight to patient assessments that is clinically correlated with other information and insights external to the AI model(s) at work. We further discussed challenges and opportunities of existing image reporting practices and data characteristics, and their implication for AI data curation. Specifically, we: (i) drew attention to image quality and patient factors that can bias model training and affect image interpretation for NGT placement; (ii) described potential trade-offs between efforts to achieve high-quality training data for producing reliable AI outputs and consequential data exclusion for the potential reach of a resulting AI application; and (iii) discussed difficulties in achieving robust ground truth data labels for more borderline NGT cases. We suggested that standardized reporting templates may help to define the AI model's capabilities as well as evaluations for the NGT verification task. 

Whilst our research centred on the specific use case of NGT verification, the insights provided translate well to other medical lines and tubes investigations (e.g., CVCs, ETTs), as well as clinically relevant or critical findings detection within radiology imaging. More broadly, our leanings guide the future design and development of AI applications that are clinically useful, ethical, and acceptable in real-world healthcare services.

\section{Acknowledgments}
Special thanks go to all our study participants for their time and invaluable input to the research. JJ was supported by the Wellcome Trust [209553/Z/17/Z] and the NIHR UCLH Biomedical Research Centre, UK.

%% The next two lines define the bibliography style to be used, and
%% the bibliography file.
\bibliographystyle{ACM-Reference-Format}
\bibliography{sample-base}

%%
%% If your work has an appendix, this is the place to put it.
\appendix

\section{Appendix}
\bgroup
\def\arraystretch{1.8}% 
\begin{table*}

  \caption{Overview of existing ML/ AI work for lines and tube detection from Chest X-ray images.(*) Indicates studies with neonatal~\cite{henderson2021automatic} or pediatric~\cite{kao2015automated} populations.}
  \label{Table: TubeStudies}
 
  %\small
  \fontsize{6}{7}\selectfont
  \begin{tabular}{p{0.15\textwidth} p{0.2\textwidth} p{0.3\textwidth} p{0.3\textwidth}}
    \toprule
    \textbf{Reference} & \textbf{Research Aim }& \textbf{Dataset} & \textbf{AI/ ML Outcome}\\
    \midrule
    Abbas et al.~\cite{abbas2022automatic} & Detect presence of tube type \& Classify position of ETTs, CVCs, NGTs and Swan Ganz as normal, borderline or abnormal. & RANZCR CLiP~\cite{tang2021clip} & Use of transfer learning via EfficientNet (B7 with Auxiliary connection) achieved average AUC of 0.963; the authors also experimented with Quantization as a technique to increase inference speed and downsize model weights. \\
    
    Aryal \& Yahyasoltani ~\cite{aryal2021identifying} & Classify position of ETTs, CVCs, NGTs and Swan Ganz as normal, borderline or abnormal. & RANZCR CLiP~\cite{tang2021clip}: 9085 (out of 30083) annotated for catheter &  Use of GANs to expand dataset catheter annotations significantly improves classification accuracy from AUC of 0.87 (CNN without synthetic annotations) to 0.96.\\  
    
    Borvornvitchotikarn \& Yooyativong~\cite{borvornvitchotikarn2022pre} & Classify position of ETTs, CVCs, NGTs and Swan Ganz as normal, borderline or abnormal. & RANZCR CLiP~\cite{tang2021clip}: 24,062 for training; 6,021 for validation; 3,255 for testing & Use of novel spatial attention module (called Pac-SA) based on an attention mechanism to enhance multi-label image classification. Best performing model achieves 94.65\% Accuracy and 65.18\% Precision.\\

    Drodov et al.~\cite{ drozdov2023artificial} & Detect NGT malposition on CXRs \& evaluate model impact as clinical decision support tool. & CXRs from 14 acute sites in NHS Greater Glasgow and Clyde~\cite{ PublicHealthScotland}: 1,132,142 CXRs pre-training (ImageNet); 7,081 CXRs fine-tuning; 335 CXRs evaluation & Their model ensemble achieved classification AUCs of 0.82 for satisfactory; 0.77 for malpositioned; and 0.98 for mispositioned into the lungs.\\

    Elaanba et al.~\cite{elaanba2021automatic} & Detect (binary) position labels for ETTs, CVCs, NGTs and Swan Ganz. & RANZCR CLiP~\cite{tang2021clip} & Their best performing model in this multilabel classification tasks achieved an AUC of 80\%.\\
    
    Hendersen et al.~\cite{henderson2021automatic} (*) & Detect presence and tube type: ETTs, NGTs, umbilical arterial and venous catheters (UACs, UVCs). & 777 neonatal AP chest + abdominal radiographs (49 no catheter; 167 with 1 catheter; 561 with 2+ catheters); labelled by medical student, reviewed by a resident and attending paediatric radiologist; Obtained from NICU of Royal University Hospital, Saskatoon, Saskatchewan, CA (2014-2015). & Average precision achieved in detecting the presence of NGTs was 97-99\%; 98-99\% for ETTs, 93-98\% for UACs, and 93\% for UVCs; authors also analysed performance based upon number of catheters in each image.\\

    Kao et al.~\cite{kao2015automated} (*) & Detect presence of ET tube + tip location in paediatric CXRs. & 528 CXRs with ETTs + 816 CXRs without ETTs from 412 patients in NICU of Kaohsiung Medical University Hospital, Taiwan (Jan-July 2013). & AUC of 94.3\% in detecting existence of an ETT with a tip location detection error of 1.89 ± 2.01 mm.\\

    Khan \& Ali~\cite{khan2021early} & Classify ETTs, CVCs, NGTs and Swan Ganz as present, normal, borderline or  abnormal. & RANZCR CLiP~\cite{tang2021clip}: 9083 segmented CXRs NIH ChestX-ray14~\cite{wang2017chestx}. & Use of UNet for segmentation and  transfer learning via EfficientNet for classification; achieving an AUC score of 0.972.\\

    Lee et al.~\cite{lee2018deep} & Detect PICC line tip location. & 600 de-identified, HIPPA compliant DICOM AP CXRs images from 600 patients with visible PICCs; obtained from Massachusetts General Hospital, USA (Jan 2015- Jan 2016). & Best model obtained absolute distances from ground truth with a mean of 3.10 mm, a standard deviation of 2.03 mm, and a root mean squares error (RMSE) of 3.71 mm.\\

   Rungta~\cite{rungta2021detection} & Classify position of ETTs, CVCs, NGTs and Swan Ganz as normal, borderline or abnormal. & RANZCR CLiP~\cite{tang2021clip} & Report EfficientNet accuracy of 0.89 for validation and 0.91 for test datasets. \\
    
   Seah et al.~\cite{seah2021effect} & Classify 127 clinical findings from chest X-ray, which included 34 crucial clinical findings (e.g., suboptimal tube placement) \& evaluate model impact as clinical decision support tool. & 5 Datasets: MIMIC~\cite{ johnson2019mimic}, I-MED, NIH ChestX-ray14~\cite{ wang2017chestx}, CheXpert~\cite{ irvin2019chexpert}, PadChest~\cite{ bustos2020padchest}: 821,681 CXR images for model training; 2,589 enriched CXRs for testing. & Report AUCs of their DL model for  suboptimal: NGT (0.984), ETT (0.995), Central line (0.969), and pulmonary arterial catheter (0.992).\\

    Singh et al.~\cite{singh2019assessment} & Detect critical vs. non-critical placement of enteric feeding tubes. & 5475 de-identified HIPAA compliant chest + abdominal X-rays (5301 non-critical); 2 expert labellers; data source unknown. & Best performing models achieved AUCs of 0.82-0.85 in differentiating critical vs. non-critical placement. \\

    Sirazitdinov et al.~\cite{sirazitdinov2021landmark} & Detect malpositioned CVC tip utilizing 13 distinct anatomical landmarks . & NIH ChestX-ray14 ~\cite{ wang2017chestx}: 300 manually selected CXRs that had CVCs present; annotated 13 landmarks + CVC tip position. & Model achieved an AUC of 0.96 for identifying malpositioned CVC tips.\\

    Sreedhar et al.~\cite{sreedhar2021detection} & Detect the presence and correct placement of multiple catheters. & RANZCR CLiP~\cite{tang2021clip} & Proposal of using Mask R-CNN technique.\\

    Subramanian et al.~\cite{subramanian2019automated} & Detect presence and differentiate between four CVC types (PICCs, Swan-Ganz, IJ lines, Subclavian lines). & NIH ChestX-ray14 ~\cite{ wang2017chestx}: 1500 AP CXRs (608 pixel-level CVC type annotations); 3000 CXR (2381 with external medical device, image-level); 10,746 CXRs with at least 1 CVC (Image-level). & Best performing models achieved 85.2\% accuracy in CVC detection (91.6\% precision) and CVC type classification (95.2\% precision).\\

    Yu et al.~\cite{yu2020detection} & Segment PICC line and detect its tip. & 348 AP CXRs from 326 patients with visible PICCs for segmentation tasks; a subset (174 CXR) labelled for tip detection task; obtained from Quhua Hospital of Zhejiang Province and Hospital of Zhejiang University. & Their model on catheter segmentation achieved an F1 score of 0.58 on both the test and validation set; and an F1 score of 0.74 on the test and 0.69 on the validation set for tip detection.\\

    \bottomrule
  \end{tabular}
\end{table*}
%\subsection{Part One}
%X

%\subsection{Part Two}
%X

%\section{Online Resources}
%X

\end{document}